\documentclass[12pt,pre,aps,superscriptaddress]{revtex4}
\usepackage[T1]{fontenc}
\usepackage{graphicx}
\newcommand{\beq}[2]{\begin{equation}#1\label{#2}\end{equation}}
\newcommand{\ceq}[1]{(\ref{#1})}
\newcommand{\mbd}[1]{\mbox{\bf #1}}
\newcommand{\bp}{\mbd{p}}

\newcommand{\bs}{\mbd{s}}
\newcommand{\br}{\mbd{r}}
\newcommand{\bx}{\mbd{x}}
\newcommand{\by}{\mbd{y}}
\newcommand{\bz}{\mbd{z}}
\newcommand{\ba}{\mbd{a}}

\newcommand{\bA}{\mbd{A}}

\newcommand{\bR}{\mbd{R}}

\newfont{\mbld}{cmbx10 scaled 800}
\newfont{\cab}{cmsy10 scaled 1200}
\newfont{\scab}{cmsy10 scaled 1000}
\newfont{\bcall}{cmbsy10 scaled 1200}

\newcommand{\sbr}{\mbld{r}}
\newcommand{\sbp}{\mbox{\mbld{p}}}
\newcommand{\sbx}{\mbox{\mbld{x}}}
\newcommand{\sby}{\mbox{\mbld{y}}}

\newcommand{\BF}[1]{\mbox{\boldmath $#1$}}          
\newcommand{\nablab}{\BF{\nabla}}
\begin{document}
\title{Directed
  Polymers with Constrained Winding Angle}
\author{Franco Ferrari}
\email{ferrari@univ.szczecin.pl}
\affiliation{Institute of Physics, University of Szczecin,
  ul. Wielkopolska 15, 70-451 Szczecin, Poland}
\author{Vakhtang G. Rostiashvili}\email{rostiash@mpip-mainz.mpg.de}
\author{Thomas A. Vilgis} 
\email{vilgis@mpip-mainz.mpg.de}
\affiliation{Max Planck Institute for Polymer Research, 10
  Ackermannweg, 55128 Mainz, Germany}

\begin{abstract}
In this article we study 
from a non-perturbative point of view
the entanglement of two directed
polymers subjected to repulsive interactions given 
by a Dirac $\delta-$function potential.
An exact formula of the so-called
second moment of the winding angle is derived. This result is used to
provide a thorough analysis of entanglement phenomena in the
classical system of two polymers subjected to repulsive interactions
and related problems. No approximation is made in treating the
constraint on the winding angle and the
repulsive forces. In particular, we
investigate how repulsive
forces influence the entanglement degree of the two-polymer system.
In the limit of ideal polymers, in which the interactions
are switched off, we show that
our results are in agreement with those of previous works.
\end{abstract}
\maketitle
\section{Introduction}\label{sec:intro}
The statistical mechanics of two polymers with constraints on their
winding angle
has been extensively studied in order to understand
the
behavior of physical
polymer systems, like for instance biological macromolecules of DNA
\cite{waslev} or liquid crystals composed of stacks of disk-shaped
molecules \cite{dsmol},
 see  Refs.~\cite{beli,pityor,comtet, nechaev,drossel,tanaka,nelson, moka,
khvi,feklla, otto,brescia,edwards,spitzer,prager,saito}. A detailed
 review on the subject, together with interesting proposals  of
 how to include in the treatment of topologically entangled polymer
 link invariants which are more sophisticated than the winding number,
 can be found in \cite{kleinert}.
Up to now, however, despite many efforts,
mainly ideal polymer chains or loops winding around each
others have been considered, while the repulsive interactions between the
monomers have been treated approximatively or
exploiting in a clever way
scaling arguments integrated by  numerical simulations,
as for instance in \cite{drossel}.

Here we concentrate ourselves on the case of two directed
polymers interacting via a  repulsive Dirac $\delta-$function
potential \cite{karzha,dirpoltheo}. 
We are particularly interested in the average
degree of entanglement of the system, which we wish to estimate
by computing the
square average winding angle of 
the two polymers. This quantity is
also called second moment of the winding angle or simply second
moment and is a special example of the topological moments first
introduced in Ref.~\cite{iwakim}.
To achieve our goals, we develop
 an approach, which combines quantum mechanical and
field theoretical techniques. With respect to previous works, we are
able to obtain 
exact results even if the repulsive
interactions are not switched off.

In principle, the average of
any observable like the squared winding angle can be  derived once the
partition function of the system is 
known, but in our case it turns out that the partition function is
simply too complicated to obtain any analytical result.
This happens essentially because the full 
$\delta-$function potential is not a central potential, since it mixes
both radial 
and angular variables. For this reason, the usual procedure of going to
polar coordinates and then solving the differential equation satisfied
by the partition
function of the entangled polymers with the
method of separation of variables  \cite{kleinert}, does no longer
produce simple 
formulas as in the situations in which only central forces are present.

To avoid these difficulties, one possibility is to approximate the
$\delta-$function 
potential with some radial potential, like for instance the hard core
potential of Ref.~\cite{drossel}. However, here we shall
adopt a different strategy, based on field theories, which does not
require any approximation.
This strategy has been
developed in
\cite{tanaka, feklla}
(see also Ref.~\cite{kleinert} for more
details)
to cope with ideal closed polymers whose
trajectories are concatenated. Also such systems are characterized by
a non-central
potential, which comes out as a consequence of the topological constraints
imposed on the trajectories.
In the field theoretical formulation  of Refs.~\cite{tanaka,feklla,kleinert}
the computation of the second
moment is reduced to the problem of computing some
correlation functions
of a field theory. A bonus is provided by the fact that this
computation requires just a finite number of Feynman diagrams to be
evaluated.
In the present case, due to the
presence of the $\delta-$function potential,
the field theory which we obtain is no longer free 
as that of Refs.~\cite{tanaka,feklla,kleinert}. Nevertheless, we will see that
the theory is still
linear and thus it can be exactly solved once its propagator is known.
Luckily, this propagator may be computed exactly
using powerful non-perturbative
techniques developed in the context of quantum
mechanics to deal with Hamiltonian containing
 $\delta-$function potentials, see Refs.~\cite{bethe, Faddeev, albev,
  gerbert, hagen, grosche,Park,fey,shul,jack}.
Basically, starting from the Green function of a
particle whose dynamics is governed by a given Hamiltonian ${\cal
  H}_0$, these techniques 
provide an algorithm to construct the Green function of a
particle corresponding to a {\it perturbed} 
Hamiltonian ${\cal H}={\cal H}_0+V_\delta$, where $V_\delta$ is the
$\delta-$function potential. 
One advantage of these methods is that
there is a long list of potentials
for which the Green functions of the
unperturbed Hamiltonians ${\cal H}_0$  are known. 
In this way, it is easy to generalize our treatment including
new
interactions,  which could be relevant in
polymer physics, like for instance the Coulomb interaction. 
The price to be paid  is that the
quantum-mechanical algorithm
works when the Green functions are expressed as functions of the
energy instead of the time. 
In the polymer analogy, 
assuming that the ends of the polymers are attached to two planar surfaces 
perpendicular to the $z-$axis and located
at the positions
$z=0$ and $z=L$,
the role of time is played by the distance $L$,
while the energy  corresponds to the chemical potential conjugated $L$. 
To recover the original dependence on $L$,  one needs
to calculate an inverse Laplace transform of the field
propagator with respect to the energy. In general, this is not
a simple task. 

Once the
propagator of the linear field theory is known, the
correlation functions which enter in the expression of
the second moment may be calculated contracting the
fields in all possible ways using the Wick prescription. 
 At the
end, we get in this way an exact formula of the second moment
as a function of the {\it energy}, which, we remember, has here the
meaning of the chemical potential conjugated to the distance $L$.
In the $L$ space, 
due to the problems of computing  the
inverse Laplace transform of the propagator mentioned above,
only an approximated expression of the
second moment 
will be given in the limit of large values of $L$ and assuming that
the strength of the $\delta-$function potential is weak enough to
allow a perturbative approach.

Our results allow both a qualitative and quantitative understanding of
the way in which the 
repulsive interactions affect the  entanglement of two
directed polymers.
The corrections introduced by these interactions
in the expression of the second moment of ideal polymers
have been studied in some interesting limits.
First of all, it has been examined
the limit of long polymer trajectories, in which we show that
repulsive interaction become particularly relevant.
Moreover, we have investigated also 
the perturbative regime and the strong coupling limit,
which is important to recover the excluded volume interactions.
While it is not a
problem to take the strong coupling limit within
our exact treatment of the repulsive interactions,
it turns out that, in this case, the expression of the second moment
is particularly complicated from
the analytical point of view. For this reason,
in the Conclusions we will discuss the application of
a powerful perturbative method to study field theories at strong
coupling
due to Kleinert \cite{klI,klII,klbookII}.
Finally, the consistency of
our results with the previous ones has been checked by studying the limit of
ideal polymers.

The material presented in this paper is divided as follows.
In the next Section, the problem of computing the second moment of the
winding angle of two directed polymers 
interacting via a $\delta-$function potential
is briefly illustrated using the path
integral approach. A constraint on the winding angle is imposed
by coupling the trajectories of the polymers with
a suitable external magnetic field,
following the strategy of 
previous works like for instance \cite{edwards,brescia,tanaka,kleinert}.
 In Section \ref{sec:three}  the second moment is expressed in the
 form of a
 finite sum of 
amplitudes of a linear field theory.
These amplitudes may be computed once the propagator of the
theory is constructed.
In our case, the propagator coincides with the Green function 
of a particle diffusing in a $\delta-$function potential. The
derivation of this 
Green function in the
energy representation 
using
non-perturbative 
techniques developed in the context of quantum mechanics
\cite{bethe, Faddeev, albev, gerbert,
  hagen, grosche, Park,fey,shul,jack}
is the subject of Section \ref{sec:four}. 
The $\delta-$function potential is
responsible of the appearance of singularities in the propagator at
short distances, which have been regulated here with the introduction of
a cut-off. This procedure is
 motivated by the fact that in polymer physics there is no
point in considering distances which are smaller than the dimensions
of a monomer. A comparison with the more rigorous method of
renormalization is made, showing the consistency of the two
procedures.
The propagator derived in Section~\ceq{sec:four} has a particularly
nice form, in which the contributions coming from the repulsive forces
can be separated from the free part of the propagator, which
is related to the random walk of ideal polymers.
This splitting of the propagator is used in
Section \ref{sec:five} to discuss
qualitatively and qualitatively the effects of the $\delta-$function
interactions on the entanglement of the system. 
The results of Sections \ref{sec:three} and \ref{sec:four}
 provide in principle all the ingredients of
the second moment. However, the
amplitudes of the linear field
theory derived in Section~\ref{sec:three} should still be evaluated.
In this task one encounters the typical problems 
occurring in the evaluation of the analytical expressions of Feynman
diagrams. 
In the case of the second moment there are just tree diagrams, but
still one has to perform complicated
integrations 
over the spatial coordinates
which are transverse to the $z-$axis.
Even assuming that polymers are ideal,
the analytical evaluation of these integrations requires drastic
approximations, see for instance \cite{tanaka}.
To avoid these difficulties,
we average
 the second moment 
with respect to the positions of the endpoints of the two polymers.
This averaged version of the second moment can be computed
without any approximation in the energy representation. This is done
in Section \ref{sec:six}.
The expression of
the averaged second moment in the
$L-$space is provided  instead only at the first  perturbative order
in the strength of the repulsive potential
and assuming additionally  that the value of $L$ is large.
We give also an exact formula of the second moment without performing
any averaging procedure
as a function of
$L$. This formula is however explicit only up to the calculation of
the inverse Laplace 
transform of the propagator derived in Section \ref{sec:four}.
In  Section \ref{sec:seven}
we consider  the situation in which
the polymers are not interacting in order to allow the comparison with
previous results. 
Finally, the discussion of the obtained results and ideas for further
developments are presented in the Conclusions.

\section{The Statistical Mechanics of Two Directed Polymers with
  Constrained Winding angle}\label{sec:two}
Our starting point is the action of two directed polymers $A$ and $B$:
\beq{{\cal A}_0=\int_0^L
dz\left[ c\left(\frac{d\br_A}{dz}\right)^2+
c\left(\frac{d\br_B}{dz}\right)^2-V(\br_A-\br_B)\right]
}{actnorm}
where $V(\br_A-\br_B)$ is the  potential:
\beq{
V(\br_A-\br_B)=-v_0\delta(\br_A-\br_B) \qquad\qquad v_0>0
}{dirpot}
The sign of $v_0$ has been chosen in such a way that the interaction
associated to the potential $V(\br)$
is repulsive.
The parameters $c$ and $L$ 
determine
the average length of the
trajectories of the polymers.
%
 The ends of the polymers are
supposed to be fixed on two surfaces perpendicular to the $z-$axis and
located at the heights $z=0$ and $z=L$. 
Both polymers have a preferred direction along the $z$ direction.
The vectors
$\br_A(z)$ and $\br_B(z)$, $0\le z\le L$, measure the polymer displacement
along the remaining two directions of the space. 

The action of Eq.~\ceq{actnorm} resembles that of two quantum
particles in the case of imaginary time $z$. To stress these analogies with
quantum mechanics, the $z-$variable will be treated as a pseudo-time
and renamed using from now on the letter $t$ instead of $z$.

In the system of the center of mass:
\beq{\br=\br_A-\br_B\qquad\qquad\bR=\frac{\br_A+\br_B}2}{cmcoor}
the action \ceq{actnorm} becomes:
\beq{{\cal A}_0=\int_0^Ldt\left[
\frac c2\left(
\frac{d\br}{dt}\right)^2+2c\left(\frac{d\bR}{dt}\right)^2-V(\br)\right]
}{cmact}
The motion of the center of mass, which is a free motion
described by the coordinate $\bR(t)$, will be ignored.

We consider the partition function of the above two-polymer system
with the addition of a constraint on the entanglement of the trajectories:
\beq{
{\cal Z}_m=\int{\cal D}\br e^{-\int_0^Ldt\left[
\frac c2\left(\frac{d\sbr}{dt}\right)+V(\sbr)\right]
}
\delta(m-\chi)
}{totparfun}
$\chi$ is the so-called winding angle. Its expression is given by:
\beq{\chi=\int_0^L\bA(\br(t))\cdot d\br(t)}{wnum}
where  $\bA(\br)$ is a vector potential with components:
\beq{
A_j(\br)=\frac 1{2\pi}\epsilon_{ij}\frac{x^i}{\br^2}\qquad\qquad i,j=1,2
}{vecpot}
In the above equation  we have represented the vector $\br$
using cartesian coordinates $x^1,x^2$. i.~e.
 $\br=(x^1,x^2)$. Moreover, from now on, middle latin indices
$i,j,\ldots=1,2$ will label the directions which are perpendicular to the
$t-$axis.
The definition of the partition function ${\cal Z}_m$ is completed by
the boundary conditions at $t=0$ and $t=L$:
\beq{\br(0)=\br_0\qquad\qquad\br(L)=\br_1}{bconds}
The quantity in Eq.~\ceq{wnum} becomes a topological invariant if
 the polymer trajectories are closed. In the present case, in which
 the trajectories are open, $\chi$ just
counts the angle with which one polymer winds up around the other.
Thus, the partition function ${\cal Z}_m$ gives the formation
probability of 
polymer paths winding up of an angle
\beq{\Delta\theta=2\pi m}{diffang}

Exploiting the Fourier representation of Dirac $\delta-$functions
\beq{
\delta(m-\chi)=\int_{-\infty}^{+\infty}\frac{d\lambda}{2\pi}
e^{i\lambda(m\chi)} 
}{fourepddf}
Eq.~\ceq{totparfun} can be rewritten as follows:
\beq{{\cal Z}_m
=\int_{-\infty}^{+\infty}\frac{d\lambda}{2\pi}e^{im\lambda}{\cal Z}_\lambda
}{ftparfun}
where
\beq{
{\cal Z}_\lambda=\int{\cal D}\br e^{-\int_0^Ldt{\cal L}}
}{afterft}
The Lagrangian ${\cal L}$ is that of a particle immersed in the
magnetic potential associated to the vector field \ceq{vecpot}:
\beq{
{\cal L}=\frac
c2\left(\frac{d\br}{dt}\right)^2+i\lambda\frac{d\br}{dt}\cdot \bA-V(\br)
}{newaction}
The Fourier transformed partition function ${\cal Z}_\lambda$  is
the grand canonical 
version of the original partition function ${\cal Z}_m$, in which the
number $m$ is allowed to take any possible value.

${\cal Z}_\lambda$ coincides with the
propagator ${\cal G}_\lambda(L;\br_1,\br_0)$, which satisfies the following
pseudo-Schr\"odinger equation:
\beq{
\left[\frac{\partial}{\partial L}-{\cal
      H}\right]{\cal G}_\lambda(L;\br_1,\br_0) =0
}{pseudseq}
${\cal H}$ is the Hamiltonian of the system, computed starting from the
Lagrangian \ceq{newaction}:
\beq{{\cal H}=\frac1{2c}(\nablab-i\lambda\bA)^2+V(\br)}{ham}
Eq.~\ceq{pseudseq} is completed by the boundary condition at $L=0$:
\beq{{\cal G}_\lambda(0;\br_1,\br_0)=\delta(\br_1-\br_0)}{bbdc}


The
 average degree of entanglement of the two polymers can be
estimated computing the topological moments of the winding angle
 $\langle m^{2k}\rangle_{\br_1,\br_0}$, 
$k=0,1,2,\ldots$ \cite{iwakim}. Once the partition function is
 known, the  $\langle m^{2k}\rangle_{\br_1,\br_0}$ may be expressed as follows:
\beq{
\langle m^{2k}\rangle_{\br_1,\br_0}=
\frac{
\int_{-\infty}^{+\infty} dm \,m^{2k} {\cal Z}_m}
{\int_{-\infty}^{+\infty} dm \, {\cal Z}_m}
=
\frac{
\int_{-\infty}^{+\infty} dm \,m^{2k}
\int_{-\infty}^{+\infty} \frac{d\lambda}{2\pi}e^{im\lambda}{\cal
  G}_\lambda(L;\br_1,\br_0)}
{\int_{-\infty}^{+\infty} dm \,
\int_{-\infty}^{+\infty} \frac{d\lambda}{2\pi}e^{im\lambda}{\cal
  G}_\lambda(L;\br_1,\br_0)}
}
{sectopmomdef}
The quantities
 $\langle m^{2k}\rangle_{\br_1,\br_0}$ depend on the
 boundary conditions
$\br_0,\br_1$ and, of course, on the parameters $c$ and $L$.
For practical reasons, we will also  consider  the following {\it
  averaged}
topological moments:
\beq{
\langle
  m^{2k}\rangle=
\frac{
\int
d^2r_0
\int
d^2r_1 
 \int dm m^{2k}{\cal Z}_m}
{
\int
d^2r_0
\int
d^2r_1 
 \int dm {\cal Z}_m}
}
{aderr}
As Eq.~\ceq{aderr} shows, the average is
  performed 
 with respect to the relative positions 
$\br_0,\br_1$
of the endpoints. This
  is equivalent to an average over the positions of the endpoints
  $\br_A(t), \br_B(t)$ at the instants $t=0$ and $t=L$, because
the 
coordinates of the center of mass
have been factored out from the
  partition function and thus they
 do not play any role.
The advantage of the averaged topological moments is that, a
  posteriori, it will be seen that
 their computation is
 easier than that of the topological moments given in
  Eq.~\ceq{sectopmomdef}. From the physical point of view, the
  averaged topological moments measure the entanglement of two
  polymers, whose ends at the instants $t=0$ and $t=L$ are free to move.

Here we  will be
interested only in the 
second  moment $\langle m^2\rangle_{\br_1,\br_0}$
and in the averaged second moment  $\langle m^2\rangle$, i. e. in
the case $k=1$ of Eqs.~\ceq{sectopmomdef} and \ceq{aderr}.
The second moment is in fact enough in order to estimate
the formation probability of entanglement with a given winding
angle and to
determine how the winding angle grows with increasing polymer
lengths.

In the following it will be useful to work
in the so-called energy
representation, i.~.e considering the Laplace transformed of the partition
function ${\cal G}_\lambda(L;\br_1,\br_0)$ with respect to $L$:
\beq{
{\cal G}_\lambda(E;\br_1,\br_0)=
\int_0^{+\infty}dL e^{-EL}
{\cal G}_\lambda(L;\br_1,\br_0)
}{laplrep}
The new partition function ${\cal G}_\lambda(E;\br_1,\br_0)$ describes
the probability of two entangled polymers of any length 
subjected to the condition that the relative positions of the polymer
end at the initial
and final instants $t_0$ and $t_1$  are given by the
 vectors $\br_0$ and $\br_1$. With respect to the formulation in the
$L-$ space, however,
 the distance $t_1-t_0$ is no longer exactly equal to $L$, but is
 allowed to vary 
 according to a distribution which is governed by the
Boltzmann-like factor $e^{EL}$.
Thus, $E$ plays the role of the chemical potential conjugated to the
end-to-end distance of the polymer trajectories in the $t-$direction.
It is worth to remember that, roughly speaking, small values of $E$
correspond to large values of $L$, while large values of $E$
correspond to small values of $L$.
Starting from Eq.~\ceq{pseudseq} and recalling
the boundary conditions \ceq{bbdc}, it is easy to check that 
${\cal G}_\lambda(E;\br_1,\br_0)$ satisfies the stationary
pseudo-Schr\"odinger  equation:
\beq{
\left[E-{\cal H}\right]{\cal G}_\lambda(E;\br_1,\br_0)=\delta(\br_1-\br_0)
}
{pseenerep}
where ${\cal H}$ is always the Hamiltonian of Eq.~\ceq{ham}.

\section{Calculation of the second moment using the field theoretical
  formulation}\label{sec:three}
In this Section we wish to evaluate the expression of the second
moment as a function of the energy $E$ using a field theoretical
formulation of the polymer partition function. The starting point is
provided by the formula of the second moment in the $L-$space
suitably rewritten in the following way:
\beq{
\langle m^2\rangle_{\br_1,\br_0}=\frac{N(L;\br_1,\br_0)
}{
D(L;\br_1,\br_0)
}
}
{secmomnumden}
For consistency with Eq.~\ceq{sectopmomdef}, the numerator
$N(L;\br_1,\br_0)$ and the denominator $D(L;\br_1,\br_0)$ appearing in
Eq.~\ceq{secmomnumden} must be of the form:
\beq{
N(L;\br_1,\br_0)=\int_{-\infty}^{+\infty}dm\enskip m^2\int_{-\infty}^{+\infty}
\frac{d\lambda}{2\pi} e^{im\lambda}{\cal G}_\lambda(L;\br_1,\br_0)
}
{numsm}
and
\beq{
D(L;\br_1,\br_0)=\int_{-\infty}^{+\infty}dm \int_{-\infty}^{+\infty}
\frac{d\lambda}{2\pi} e^{im\lambda}{\cal G}_\lambda(L;\br_1,\br_0)
}
{densm}

Using Eq.~\ceq{laplrep}, it is now straightforward
to compute the Laplace transform of $N(L;\br_1,\br_0)$ and
$D(L;\br_1,\br_0)$:
\beq{
N(E;\br_1,\br_0)=\int_{-\infty}^{+\infty}dm m^2\int_{\infty}^{+\infty}
\frac{d\lambda}{2\pi} e^{im\lambda}{\cal G}_\lambda(E;\br_1,\br_0)
}
{momno}
\beq{
D(E;\br_1,\br_0)=\int_{-\infty}^{+\infty}dm \int_{\infty}^{+\infty}
\frac{d\lambda}{2\pi} e^{im\lambda}{\cal G}_\lambda(E;\br_1,\br_0)
}
{momnt}
Once the functions $N(E;\br_1,\br_0)$ and $D(E;\br_1,\br_0)$ are
known, one can construct the ratio:
\beq{
\langle m^2\rangle_{\br_1,\br_0}(E)=\frac{N(E;\br_1,\br_0)}{D(E;\br_1,\br_0)}
}
{secmomerep}
which is nothing but  the second moment of the
winding angle
expressed as a function of the
chemical potential $E$.

We remark that the Green function
${\cal G}_\lambda(E;\br_1,\br_0)$ 
is related to the Feynman propagator
of the spin$-\frac 12$ Aharonov-Bohm problem in quantum mechanics.
In principle, this Green function can be computed exactly
starting from Eq.~\ceq{pseenerep}
 \cite{Park}, but its
final expression is too complicated for our purposes. Moreover, the
method used in \cite{Park} to renormalize the singularities coming
from the presence of the $\delta-$function potential is valid only in
a restricted region of the domain of $\lambda$. This is incompatible
with our requirements, because, to derive the second moment, one has to
integrate ${\cal G}_\lambda(E;\br_1,\br_0)$ with respect to $\lambda$
over the whole real line.
For this reason, we prefer here to use a field theoretical
representation of this Green function. This is achieved by noting
that ${\cal
  G}_\lambda(E;\br_1,\br_0)$  coincides with the inverse matrix
element of the operator $E-{\cal H}$:
\beq{
{\cal G}_\lambda(E;\br_1,\br_0)=
\langle\br_1|\frac 1{E-{\cal H}}|\br_0\rangle}{grehj}
and may be expressed in a functional integral form in terms of replica
fields:
\beq{
{\cal G}_\lambda(E;\br_1,\br_0)=
\lim_{n\to 0}
\int{\cal D}\Psi{\cal
  D}\Psi^*\psi_1(\br_1)\psi^*_1(\br_0) e^{-S(\Psi^*,\Psi)
}}
{ftrep}
In the above equation $\Psi^*,\Psi$ are multiplets of replica fields:
\beq{\Psi=(\psi_1,\ldots,\psi_n)\qquad\qquad
\Psi^*=(\psi^*_1,\ldots,\psi^*_n)}{repfie}
with action
\beq{
S(\Psi^*,\Psi)=\int d^2x \Psi^*\star\left[
E-\frac 1{2c}(\nablab_x-i\lambda\bA)^2-v_0\delta(x)
\right]\Psi
}
{erepact}
The symbol $\star$ in Eq.~\ceq{ftrep} denotes summation over
the replica index. For example
$\Psi^*\star\Psi=\sum_{\sigma=1}^n\psi^*_\sigma \psi_\sigma$.
Below it will be used also the convention $\Psi^*\star\Psi=|\Psi|^2$.
The details of the derivation of Eq.~\ceq{ftrep} can be found in
previous publications on the subject \cite{Park,feklla} and will not
be provided here. 

In order to proceed, it will be convenient to expand the action
\ceq{erepact} in powers of $\lambda$:
\beq{
S(\Psi^*,\Psi)=S_0(\Psi^*,\Psi)+\lambda S_1(\Psi^*,\Psi)+
\lambda^2S_2(\Psi^*,\Psi) 
}{erepactdiv}
where we have put:
\beq{
S_0(\Psi^*,\Psi)=\int d^2x\left[
\frac 1{2c}|\nablab\Psi|^2+\left(
E-v_0\delta(x)
\right)|\Psi|^2
\right]
}
{szero}
\beq{
S_1(\Psi^*,\Psi)=\frac i{2c}\int d^2x\bA\cdot
\left[
\Psi^*\star(\nablab\Psi)-(\nablab\Psi^*)\star\Psi
\right]
}
{sone}
\beq{
S_2(\Psi^*,\Psi)=\frac 1{2c}\int d^2x \bA^2|\Psi|^2
}
{stwo}

At this point we come back to the computation of the quantities
$N(E;\br_1,\br_0)$ and $D(E;\br_1,\br_0)$ appearing in the expression
of the second moment.
Exploiting the new form of the partition function
given by Eqs.~(\ref{ftrep}--\ref{stwo}), together with the
relation
\beq{
\int_{-\infty}^{+\infty}dm \enskip m^\nu
e^{im\lambda}=2\pi(i)^\nu
\frac{\partial^\nu\delta(\lambda)}{\partial\lambda^\nu}
\qquad\qquad \nu=0,1,\ldots
}
{reldevs}
and the fact that ${\cal Z}_{\pm\infty}=0$, it is possible to rewrite
Eqs.~\ceq{momno} and \ceq{momnt} as follows \footnote{For details, see
  Ref.~\cite{feklla}, where a similar calculation has been done in the
  case of closed polymers.}:
\beq{
N(E;\br_1,\br_0)
=\lim_{n\to 0}\int {\cal D}\Psi^*{\cal D}\Psi
\psi_1(\br_1)
\psi_1^*(\br_0)[2S_2(\Psi^*,\Psi)-(S_1(\Psi^*,\Psi))^2]e^{-S_0(\Psi^*,\Psi) }
}{nsdo}
\beq{
D(E,\br_1,\br_0)=\lim_{n\to 0}
\int {\cal D}\Psi^*{\cal D}\Psi
\psi_1(\br_1)\psi_1^*(\br_0)e^{-S_0(\Psi^*,\Psi) }
}{nsdt}
The right hand sides of Eqs.~\ceq{nsdo} and \ceq{nsdt} represent
vacuum expectation values of a field theory governed by the action
$S_0(\Psi^*,\Psi)$ 
of Eq.~\ceq{szero}. In the formulation in terms of quantum operators we have:
\beq{
N(E;\br_1,\br_0)=\lim_{n\to 0}
\langle 0|\psi_1(\br_1)\psi^*_1(\br_0)2S_2(\Psi^*,\Psi)|0\rangle_n
-\lim_{n\to 0}
\langle 0|\psi_1(\br_1)\psi^*_1(\br_0)(S_1(\Psi^*,\Psi))^2|0\rangle_n
}{nsdoqop}
\beq{
D(E;\br_1,\br_0)=\lim_{n\to 0}
\langle 0|\psi_1(\br_1)\psi^*_1(\br_0)|0\rangle_n
}{nsdtqop}
The correlation functions have a subscript $n$ to remember that,
according to the replica method, they should be computed first
assuming that
the number of replicas $n$ is an arbitrary positive integer and then
taking the limit for $n$ going to zero.

\begin{figure}[bpht]
\centering
\includegraphics[scale=.4]{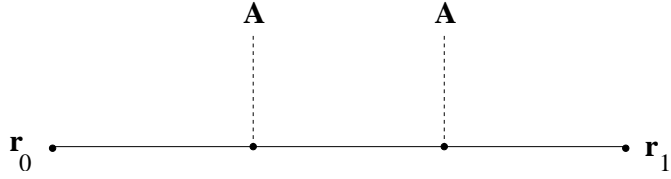}\\
\caption{Feynman diagram corresponding to the amplitudes of
  Eqs.~\ceq{amptwoone}--\ceq{amptwofour}. The two polymers $A$ and $B$
  start at a 
  distance $|\br_0|$ from each other and interact twice with the
  the external field $\mathbf A$. 
  At the end, the relative position of
  the end points at the instant $t=L$ is given by $\br_1$.
  The three-vertices appearing in the Figure
  are related to the interaction described by Eq.~\ceq{sone}.
}\label{processone}
\end{figure}

\begin{figure}[bpht]
\centering
\includegraphics[scale=.4]{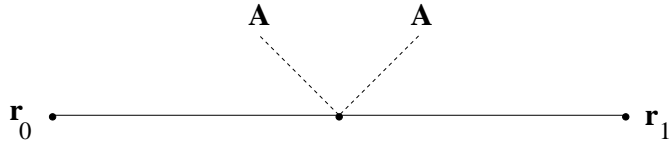}\\
\caption{Feynman diagram corresponding to the amplitude of
  Eq.~\ceq{amponek}. The two polymers $A$ and $B$ start at a
  distance $|\br_0|$ from each other and interact with the
  the external field $\mathbf A$. At the final instant $t=L$, the
  relative position of 
  the end points is given by $\br_1$.
  The four-vertex appearing in this Figure
  is related to the interaction described by Eq.~\ceq{stwo}.
}\label{processtwo}
\end{figure}

The above correlators  may be evaluated
 using
standard field theoretical methods. One could be tempted to use a
 perturbative approach assuming that the value of $v_0$ appearing in
the action $S_0(\Psi^*,\Psi)$ 
of Eqs.~\ceq{nsdo} and \ceq{nsdt} is
 small,
but this is not necessary. As a matter of fact,
if it is true that $S_0(\Psi^*,\Psi)$ 
does not describe
free fields
because of the presence of the
$\delta-$function potential, it is also true that
it is just quadratic in the fields.
As a consequence, one is allowed to define a
propagator $G(E;\bx,\by)$ associated with this action.
It is easy to check that $G(E;\bx,\by)$  satisfies
 the equation:
\beq{
\left[
E-\frac1{2c}\nablab_{\mathbf x}^2-v_0\delta(\bx)
\right]G(E;\bx,\by)=\delta(\bx,\by)
}{fdfjh}
Using the above propagator, one can evaluate the amplitudes in
Eqs.~\ceq{nsdoqop} and \ceq{nsdtqop} exactly
by contracting the fields in all
possible ways according to the Wick theorem. After straightforward
calculations, one finds:
\beq{
\lim_{n\to 0}
\langle 0|\psi_1(\br_1)\psi^*_1(\br_0)|0\rangle_n=G(E;\br_1,\br_0)
}
{ampzero}
\beq{
\lim_{n\to 0}
\langle 0|\psi_1(\br_1)\psi^*_1(\br_0)S_2(\Psi^*,\Psi)|0\rangle_n=K(\br_1,\br_0)
}
{ampone}
\beq{
\lim_{n\to 0}
\langle
0|\psi_1(\br_1)\psi^*(\br_0)(S_1(\Psi^*,\Psi))^2|0\rangle_n=
I_1(\br_1,\br_0)+I_2(\br_1,\br_0)+I_3(\br_1,\br_0)+I_4(\br_1,\br_0)}
{amptwo}
where
\begin{eqnarray}
K(\br_1,\br_0)&=&\frac 1{2c}\int
d^2x\bA^2(\bx)G(E;\br_1,\bx)G(E;\bx,\br_0)\label{amponek} \\
I_1(\br_1,\br_0)&=&-\frac1{2c^2}\int d^2xd^2y
\left[A_i(\bx)G(E;\bx,\br_1)(\nabla_{\mathbf x}^iG(E;\by,\bx))A_j(\by)
(\nabla_{\mathbf y}^jG(E;\br_0,\by) )\right]\label{amptwoone}\\
I_2(\br_1,\br_0)
&=&+\frac1{2c^2}\int d^2xd^2y\left[
A_i(\bx)(\nabla_{\mathbf x}^iG(E;\bx,\br_1))G(E;\by,\bx)
A_j(\by)(\nabla_{\mathbf y}^jG(E;\br_0,\by))\right]\label{amptwotwo}\\  
I_3(\br_1,\br_0)&=&
+\frac1{2c^2}\int d^2xd^2y\left[
A_i(\bx)G(E;\br_0,\bx)(\nabla_{\mathbf x}^i
\nabla_{\mathbf y}^jG(E;\bx,\by))A_j(\by)G(E;\by,\br_1)\right]\label{amptwothree}\\
I_4(\br_1,\br_0)
&=&
-\frac1{2c^2}\int d^2xd^2y
\left[
A_i(\bx)G(E;\br_0,\bx)(\nabla_{\mathbf x}^iG(E;\bx,\by))
A_j(\by)(\nabla_{\mathbf y}^jG(E;\by,\br_1))\right]\label{amptwofour}
\end{eqnarray}
From the physical point of view, the above equations may be
interpreted in the following way. The fields $\Psi(\bx)$
and $\Psi^*(\mathbf x)$ contain operators which, inside each replica
sector,
 create and annihilate
 segments of the two polymers, whose relative positions
are given by the vector $\bx$. 
The two
polymer system has been projected in the two-dimensional plane
perpendicular to the $t-$axis. For this reason,  there appear only the
transverse coordinates $\bx$. The only remnant of the third
dimension is the dependence on the energy $E$. 
The correlation functions \ceq{ampone} and
\ceq{amptwo} describe the fluctuations of the two polymers
 immersed in
the $\delta-$function potential and subjected to the
interactions represented by the vector potential \ceq{vecpot}.
We recall that the origin of the latter interactions is the
presence of the 
constraint 
on the winding angle in the partition function \ceq{totparfun}.
To evaluate the correlation functions \ceq{ampone} and
\ceq{amptwo}, one needs to consider only a finite
number of Feynman diagrams, corresponding to the relevant processes
with which the two polymers interact together.
The result, after the analytical evaluation of these diagrams, is
provided by Eqs.~(\ref{amptwoone}--\ref{amptwofour}).
Let us note that in these equations the repulsive interactions due to
the $\delta-$function 
potential are
hidden in the propagators $G(E;\bx,\by)$.
The Feynman diagrams related to the amplitudes of
Eqs.~\ceq{amptwoone}--\ceq{amptwofour} are all
 topologically equivalent to the diagram of
Fig.~\ref{processone}.
The amplitude of Eq.~\ceq{amponek} is related instead to the Feynman diagram
of Fig.~\ref{processtwo}.
The vectors $\mathbf
r_1$ and $\mathbf r_0$ denote the relative positions of the end
points of the two polymers at the initial and final instants, as
already mentioned. The integration variables $\mathbf x$ and $\mathbf
y$ appearing in Eqs.~(\ref{amponek}--\ref{amptwofour}) may be regarded
as the vectors which give the
relative positions of the trajectories of the two
polymers at the instants in which they interact together via the external
vector potential
$\bA$ of Eq.~\ceq{vecpot}.  
There is no restriction on the domain of
integration of $\mathbf x$ and $\mathbf
y$, so that the components of these 
relative vectors are allowed to take any value. 
This implies that the
distance between the  polymer
segments when the interaction with $\bA$ occurs
can be arbitrarily large.

Now that the correlation functions which are present in the
 expressions of
$N(E;\mathbf
r_1,\mathbf r_0)$ and 
$D(E;\mathbf
r_1,\mathbf r_0)$ 
given in Eqs.~\ceq{nsdoqop} and \ceq{nsdtqop} have been evaluated,
see
  Eqs.~(\ref{amponek}--\ref{amptwofour}), we may put everything together
and give to the second moment of
Eq.~\ceq{secmomerep} a more explicit form:
\beq{
\langle m^2\rangle_{\mathbf r_1,\mathbf r_0}=
\frac{
2K(\mathbf r_0,\mathbf r_0)-\sum_{\omega=1}^4I_\omega(\mathbf r_0,\mathbf r_0)
}{G(E;\mathbf r_0,\mathbf r_0)}
}
{feesm}
In conclusion, the initial problem of computing the second moment of
the winding angle $\langle m^2\rangle_{\br_1,\br_0}$ has been reduced
to the evaluation of a finite number of integrals, which are
given in Eqs.~(\ref{amponek}--\ref{amptwofour}).
Of course, to make these integrals really explicit,
we still need to derive the propagator $G(E;\bx,\by)$, which is so far
the only
missing ingredient. This will be done
in the next Section.
\section{Green Functions in the Case of Hamiltonians with a
  $\delta-$function Potential}\label{sec:four}
Let $G_0(L;\bx,\by)$
be the solution of the differential equation:
\beq{
\left(
\frac\partial{\partial L}-{\cal H}_0
\right)G_0(L;\bx,\by)=0
}
{diffexac}
for a given Hamiltonian ${\cal H}_0$. When $L=0$,
 $G_0(L;\bx,\by)$ satisfies
 the boundary condition:
\beq{
G_0(0;\bx,\by)=\delta(\bx-\by)
}{bdfd}
In the case of a Hamiltonian ${\cal H}$, obtained by adding to ${\cal
  H}_0$ a $\delta-$function potential as a perturbation:
\beq{{\cal H}(\bx)={\cal H}_0(\bx)-v_0\delta(\bx)}{pertham}
we consider the analogous differential problem:
\beq{
\left(
\frac\partial{\partial L}-{\cal H}
\right)G(L;\bx,\by)=0
}
{diffcomp}
\beq{
G(0;\bx,\by)=\delta(\bx-\by)
}{bdnfd}
We wish to compute $G(L;\bx,\by)$ 
starting from the Green function
$G_0(L;\bx,\by)$, which is supposed to be known.
It is possible to show that $G(L;\bx,\by)$  and $G_0(L;\bx,\by)$  are
related by the integral equation \cite{fey,shul}:
\beq{
G(L;\bx,\by)
=G_0(L;\bx,\by)-v_0\int_0^Lds\int d^2z
G_0(L-s;\bx,\bz)\delta(\bz)
G(s;\bz,\by)
}{gkk}
We see that in the right hand side of the above equation the presence
of the $\delta-$function forces us to consider the functions
$G_0(L;\bx,\by)$ and $G(L;\bx,\by)$ evaluated
at the points 
$\bx=0$ and/or $\by=0$. Usually, at these points  Green
functions may be not well defined due to the presence of
singularities. A concrete procedure to remove these 
singularities will be indicated later. For the moment, we go further
with formal manipulations, assuming that some kind of
consistent regularization of the possible divergences has been introduced.

First of all, we perform the integration over $d^2z$ in Eq.~\ceq{gkk}:
\beq{
G(L;\bx,\by)
=G_0(L;\bx,\by)-v_0\int_0^Lds
G_0(L-s;\bx,0)
G(s;0,\by)
}{beflap}
The integral in $ds$ appearing in the right hand side of
Eq.~\ceq{beflap} is a convolution which can be better treated after a
Laplace transform.
Thus, we transform
both sides of this equation with respect to
$L$:
\beq{
G(E;\bx,\by)
=G_0(E;\bx,\by)-v_0
G_0(E;\bx,0)
G(E;0,\by)
}{lapltragf}
where 
\beq{
G(E;\bx,\by)=\int_0^{+\infty}
e^{-EL}G(L;\bx,\by) dL}{dfdfd}
and 
\beq{G_0(E;\bx,\by)=\int_0^{+\infty}
e^{-EL}G_0(L;\bx,\by) dL}{dadee}
At this point, it is easy to extract from Eq.~\ceq{lapltragf}
 the expression of $G(E;\bx,\by)$:
\beq{
G(E;\bx,\by)=G_0(E;\bx,\by)-\frac{
G_0(E;\bx,0)G_0(E;0,\by)
}{\frac 1{v_0}+G_0(E;0,0)}
}{finald}
The above formula may be used in order to solve  Eq.~\ceq{fdfjh}.
In this case, ${\cal H}_0$ coincides with the free action:
\beq{{\cal H}_0=\frac 1{2c}\nablab^2}{freeham}
and the function $G_0(E;\bx,\by)$ is given by:
\beq{
G_0(E;\bx,\by)=\frac c\pi K_0(\sqrt{2Ec}|\bx-\by|)
}{gzspec}
Here $K_0(z)$ denotes the modified Bessel function of the second kind
of order zero.

Clearly, we cannot apply directly Eq.~\ceq{finald} without introducing
a regularization. As a matter of fact, if not treated,
the naive denominator in the second
term of the right hand side is equal to infinity, i. e.  ${\frac
  1{v_0}+G_0(E;0,0)}=+\infty$. This is due to the fact that $K_0(z)$ diverges
logarithmically in the limit $z\rightarrow 0$:
\beq{
K_0(z)\sim -\log z\qquad\qquad\mbox{for $z\sim 0$}
}
{fdsf}

A natural regularization is suggested by the fact that, in polymer
physics, it has no sense to consider lengths which are smaller than
the size of the molecules which compose the polymers.
Thus, it seems reasonable to regulate ultraviolet divergences by
introducing a cut-off 
$a$ at short distances. The length $a$ is comparable with the
molecular size. According to this prescription, by inserting the Green
function of Eq.~\ceq{gzspec} in Eq.~\ceq{finald}, we obtain:
\beq{
G(E;\bx,\by)\equiv\frac c\pi
K_0(\sqrt{2Ec}|\bx-\by|)-\left( 
\frac c\pi\right)^2
\frac{K_0(\sqrt{2Ec}|\bx|)K_0(\sqrt{2Ec}|\by|)
}{\frac 1{v_0}+\frac c\pi K_0(\sqrt{2Ec}a)}
}{gffin}
The symbol $\equiv$ means that the quantity in the
left hand side of an equation
has been replaced in the right hand side with its regulated version.
The above Green function is what we need in order to evaluate
explicitly the amplitudes of Eqs.~(\ref{ampzero}--\ref{amptwo}).

The infinities coming from the $\delta-$function
potential should be treated with some care in order to avoid ambiguities.
For this reason, we would like to
compare the naive 
prescription used here to derive Eq.~\ceq{gffin} with the more
rigorous procedure of renormalization. 
It is known in fact that the renormalization of  the infinities coming
from $\delta-$function
interactions produces physically sensible results \cite{jack}.
The divergences will be regulated introducing a cut-off $\Lambda$ in
the momentum space. As a consequence, it will be
convenient to express the free Green function of Eq.~\ceq{gzspec} in
momentum space. To this purpose, we use the following formula:
\beq{
K_0(m|\bx-\by|)=\frac 1{2\pi}\int
d^2p\frac{e^{i\sbp\cdot(\sbx-\sby)}}{\bp^2+m^2} 
}
{ftsmbf}
To evaluate the Green function at the singular point $\bx=\by=0$ we
need to compute the following divergent integral:
\beq{
I(m)=\frac 1{2\pi}\int \frac{d^2p}{\bp^2+m^2}
}
{divint}
Using the above cut-off prescription to eliminate the ultraviolet
singularities we get, in the assumption $\Lambda^2\gg m^2$:
\beq{
I(m)\sim\log\frac\Lambda m
}
{divinteva}
Now, according to the spirit of renormalization, we subtract the
infinities from the bare parameters of the theory.
In our case, after choosing an arbitrary mass scale $\mu$, which gives
the renormalization point, we renormalize the bare coupling constant
$v_0$.
Actually, it will be better to call it
$v_{bare}$ instead of $v_0$ in order to distinguish it from the
effective coupling constant $v_0$ appearing in Eq.~\ceq{gffin}.
The subtraction of infinities is performed in such a way that the
quantity:
\beq{
\frac 1{v_{bare}}-G_0(E;0,0)=\frac 1{v_{ren}}+
\frac c{2\pi}\log\left(
\frac{\Lambda^2}{\mu^2}
\right )-
\frac c{2\pi}\log\left(
\frac{m^2}{\mu^2}
\right )
}
{renide}
becomes finite. We choose a sort of minimal subtraction scheme, in
which the
renormalized coupling constant $v_{ren}$ is related to the bare
coupling constant $v_{bare}$ as follows:
\beq{
\frac1{v_{bare}}+\frac c{2\pi}\log\left(
\frac{\Lambda^2}{\mu^2}
\right)=\frac 1{v_{ren}}
}{barerenrel}
Applying the last two above equations back to Eq.~\ceq{finald}, we get
as a result:
\beq{
G(E;\bx,\by)=\frac c\pi K_0(\sqrt{2Ec}|\bx-\by|)-\left(
\frac c\pi\right)^2
\frac{K_0(\sqrt{2Ec}|\bx|)K_0(\sqrt{2Ec}|\by|)
}{\frac 1{v_{ren}}-\frac c{2\pi}\log\left(
\frac{2Ec}{\mu^2}
\right) }
}{gffinren}
Eqs.~\ceq{gffin} and \ceq{gffinren} are reciprocally 
compatible. In fact, since $a$
is very small, because it is the smallest possible length scale in our
polymer problem, one can use the following approximation (see
Eq.~\ceq{fdsf})
in the
denominator of the second term of Eq.~\ceq{gffin}:
\beq{
\frac 1{v_0} +\frac c\pi K_0(\sqrt{2Ec}a)\sim
\frac 1{v_0} -\frac c{2\pi} \log(2Eca)
}
{denapp}
Comparing with the analogous denominator in Eq.~\ceq{gffinren}, it is
possible to relate $a$ with 
the inverse of the mass $\mu$:
\beq{
\mu^2=\frac 1{a^2} 
}
{equicheck}
Moreover, the effective coupling constant $v_0$ of Eq.~\ceq{gffin} may be
identified with the renormalized coupling constant $v_{ren}$, which
gives the strength of the repulsive interaction \ceq{dirpot} at
distance scales of order $a$.

Before concluding this Section, we make a small digression
about the translational invariance
of the free Hamiltonian \ceq{freeham} and consequently
of the free Green function \ceq{gzspec}.
 Clearly, this is not the same translational
invariance 
that was already present in the original action \ceq{actnorm} 
due to the 
translational invariance of the potential \ceq{dirpot}.
This new invariance is
rather related to the fact that the physics of the two polymer system
in the absence of any interaction does not change if we modify the
relative positions of the polymer ends at $t=0$ and $t=L$
in a symmetric way. An example of such transformations is the
translation of both ends of polymer $A$ at the initial and final
points by a constant vector $\ba$:
\begin{eqnarray}
\br_A(0)&=&\br_A(0)+\ba\label{strazerexone}\\
\br_A(L)&=&\br_A(L)+\ba\label{straelexone}
\end{eqnarray}
As a result of the translations
(\ref{strazerexone}--\ref{straelexone}),
the relative vector $\br(t)$
of Eq.~\ceq{cmcoor} at the instants $t=0$ and $t=L$ changes as follows:
\begin{eqnarray}
\br_0'=\br_0+\ba\label{strainvzer}\\
\br_1'=\br_1+\ba\label{strainvone}
\end{eqnarray}
Clearly the propagator
\ceq{gzspec} is invariant under the above transformations.
This kind of invariance can be explained as follows. As far as the two
polymers $A$ and $B$ do not interact, each of them may be treated as
an independent system. If we translate for instance both ends of polymer $A$ at
 $t=0$ and $t=L$
in the
symmetrical way shown by Eqs.~\ceq{strazerexone} and
\ceq{straelexone}, the number of configurations of polymer $A$
and consequently the configurational entropy of the whole system
 do not
change, because the transformation is equivalent to a translation of 
polymer $A$ in the space. 
Of course, this invariance disappears as soon as the two polymers
start to interact or if they are entangled together. Indeed, if one
adds to the free Hamiltonian \ceq{freeham} a $\delta-$function
potential, the propagator stops to be translational invariant as shown
by the
Green function of Eq.~\ceq{finald}, which does not depend on the
difference $\bx-\by$.

\section{Repulsive forces and winding angles: qualitative
and quantitative considerations}\label{sec:five}
In principle we have at this point all the ingredients which are
necessary to 
compute the second moment of Eq.~\ceq{secmomerep}.
In Eqs.~\ceq{nsdoqop} and \ceq{nsdtqop}, 
in fact, the quantities
$N(E;\br_1,\br_0)$ and $D(E;\br_1,\br_0)$ are written as linear
combinations of the amplitudes of
Eqs.~(\ref{ampzero}--\ref{amptwo}), which can be explicitly
evaluated using the propagator $G(E,\mathbf u,\mathbf v)$
given in
Eq.~\ceq{gffin} 
\footnote{Throughout this Section we will use a notation
in which the vectors  $\bx,\by$ appearing in the definition of the
propagator of Eq.~\ceq{gffin} are replaced by the vectors
$\mathbf u,\mathbf v$. This is to avoid confusions with the
notation of Eqs.~(\ref{amponek}--\ref{amptwofour}), where the same pair
of vectors $\bx,\by$ has been exploited to denote dummy
integration variables.
}
and the formulas of
Eqs.~(\ref{amponek}--\ref{amptwofour}).
The remaining task is to perform the integrations over
the coordinates $\bx$ and $\by$ in
Eqs.~(\ref{amponek}--\ref{amptwofour}). 
From the analytical point of view, the evaluation of these
 integrals poses severe
technical problems, which can be solved only
with the help of drastic approximations. However, the difficulties
become milder if we average the second moment over the endpoints of
the polymers as shown in Eq.~\ceq{aderr}.
In the energy representation, which we are using, this means that we
have to consider the following averaged version of the second moment
in
Eq.~\ceq{secmomerep}:
\beq{
\langle m^2\rangle(E)=\frac{N(E)}{D(E)}
}{secmomaveerep}
where
\begin{eqnarray}
N(E)&=&\int d^2r_0\int d^2 r_1
 N(E;\br_1,\br_0)\label{ennee}\\
D(E)&=&\int d^2r_0\int d^2 r_1 D(E;\br_1,\br_0)\label{deee}
\end{eqnarray}
Accordingly, we need to integrate the quantities $K(\br_1,\br_0)$
and $I_\omega(\br_1,\br_0)$, $\omega=1,\ldots,4$ of
Eqs.~(\ref{amponek}--\ref{amptwofour}) with respect to $\br_1$ and
$\br_0$.
Putting:
\begin{eqnarray}
K(E)&=&\int d^2r_0\int d^2 r_1K(\br_1,\br_0)\label{koneee}\\
I_\omega(E)&=&\int d^2r_0\int d^2 r_1 I_\omega(\br_1,\br_0)
\qquad\qquad \omega=1,\ldots,4
\label{ionee}
\end{eqnarray}
we obtain from Eqs.~\ceq{ampone} and \ceq{amptwo} the following
expressions of $N(E)$ and $D(E)$:
\begin{eqnarray}
N(E)&=&2K(E)-\sum_{\omega=1}^4 I_\omega(E)\label{enneeexp}\\
D(E)&=&\int d^2 r_0d^2r_1G(E;\br_1,\br_0)\label{deeeexp}
\end{eqnarray}
It will also be convenient to split the propagator $G(E;\mathbf
u,\mathbf v)$ of
Eq.~\ceq{gffin} into
two contributions:
\beq{
G(E;\mathbf u,\mathbf v)=G_0(E;\mathbf u,\mathbf v)+G_1(E;\mathbf
u,\mathbf v)
}{propsplit}
where $G_0(E;\mathbf u,\mathbf v)$ is the free propagator of Eq.~\ceq{gzspec},
which is invariant with respect to the transformations
\ceq{strainvzer}
and \ceq{strainvone}, while
\beq{
G_1(E;\mathbf u,\mathbf v)=\frac c\pi\lambda(E)K_0(\sqrt{2Ec}|\mathbf u|)
K_0(\sqrt{2Ec}|\mathbf v|)
}{nontrainvpro}
In the above equation
we have isolated in the expression of $G_1(E;\mathbf u,\mathbf v)$ the
factor:
\beq{
\lambda(E)=-\frac c\pi\left(
\frac 1{v_0}+\frac c\pi K_0(\sqrt{2Ec}a)
\right)^{-1}
}
{lambe}
It is clear that the origin of the term $G_1(E;\mathbf u,\mathbf v)$ in the
propagator is due to presence of the $\delta-$function
 interaction \ceq{dirpot} in the polymer action \ceq{actnorm}.
In fact, if $v_0=0$, this term vanishes identically.
Thus, using
the splitting of the propagator of Eq.~\ceq{propsplit}, it is now
possible to separate in Eqs.~(\ref{amponek}--\ref{amptwofour}) the
contributions to entanglement given by 
repulsive forces. 

It seems natural to expand the quantities $D(E)$,
$K(E)$ and $I_\omega(E)$ defined in Eqs.~\ceq{deee},
\ceq{koneee} and \ceq{ionee}
with respect to $\lambda(E)$ as follows:
\begin{eqnarray}
D(E)&=&D^{(0)}(E)+D^{(1)}(E)\label{dexp}\\
K(E)&=&K^{(0)}(E)+K^{(1)}(E)+K^{(2)}(E)\label{expke}\\
I_\omega(E)&=&I_\omega^{(0)}(E)+I_\omega^{(1)}(E)+I_\omega^{(2)}(E)+
I_\omega^{(3)}(E)\label{expioe}
\end{eqnarray}
where the superscript
$(n)$, with $n=0,1,2,3$, denotes the order in $\lambda(E)$. There are
no higher
order terms with $n\ge 4$, so the above expansions are exact.


It is easy to show how  $K(E)$ and the
$I_\omega(E)$'s depend on the pseudo-energy $E$. After a
rescaling of the integration variables $\br_1,\br_0,\bx$ and $\by$ in
Eqs.~\ceq{ennee} and \ceq{deee}, one finds in fact that:
\beq{
K^{(n)}(E)=\lambda^n(E)E^{-2}K^{(n)}\qquad\qquad n=0,1,2
}{behaone}
\beq{
I^{(n)}_\omega(E)=\lambda^n(E)E^{-2}I^{(n)}_\omega\qquad\qquad n=0,1,2,3
}{behatwo}
where the factors $K^{(n)}$'s and the $I_\omega^{(n)}$'s are functions
of the
parameters $a$ and $c$, but not of $E$ or $v_0$. In fact,
 the coupling constant $v_0$
appears only 
inside the powers of $\lambda(E)$.
Let us note in Eqs.~\ceq{behaone} and \ceq{behatwo}
 the presence of
the overall factor $E^{-2}$  in Eqs.~\ceq{behaone} and \ceq{behatwo}.
Looking at Eq.~\ceq{enneeexp}, it is clear
that the whole function $N(E)$ is characterized
by the leading scaling behavior $N(E)\sim E^{-2}$.  In the
$L-$space, after an inverse Laplace transform, this behavior
corresponds to the following
scaling law, which is typical of ideal polymers: $N(L)\sim L$. The
powers of $\lambda(E)$,  appearing in 
the expressions of
$K^{(n)}(E)$ and $I_\omega^{(n)}(E)$, introduce corrections to this
leading behavior that are at most logarithmic in $E$. 
As a matter of fact, if the condition $2Eca^2\ll1$ is satisfied, we
have that:
\beq{
\lambda(E)\sim
-\frac c\pi\left(v_0^{-1}
-\frac c\pi\log(\sqrt{2Ec}a)
\right)^{-1}
}
{logcorr}
Naively, the above seems the only  logarithmic correction which is
possible
in
the expressions of $N(E)$ and $D(E)$ when $E$ is small. However,
that this is not true. 
In fact,
in deriving Eqs.~\ceq{behaone} and \ceq{behatwo}, we have not
considered the divergences which arise in some of the integrations
over the variables $\bx,\by,\br_0$ and $\br_1$.
After regulating these divergences with some prescription, as for
instance the ultraviolet cut-off $a$ used in Eq.~\ceq{gffin},
we will see in Section~\ref{sec:six} that the naive rescaling of
variables exploited in order 
to obtain  Eqs.~\ceq{behaone} and \ceq{behatwo} does no longer work
and one should add extra logarithmic factors to these equations.

Eqs.~\ceq{behaone} and \ceq{behatwo} may be also useful to study the
case of polymers in confined geometries. As a matter of fact, for
large values of $E$, one recovers the limit of small values of $L$, in
which the region between the initial
and final height is very narrow.
Looking at Eqs.~\ceq{behaone} and \ceq{behatwo}, it is clear that the
only interesting corrections when $E$ is large come from the powers of
$\lambda(E)$. To evaluate these corrections, one should note that
the modified Bessel function of the second kind $K_0(z)$
goes very fast to zero for large values of $z$. As a consequence,
already in the domain of 
parameters in which $2Eca^2\ge 10$, it is possible to make the
very interesting approximation 
\beq{\lambda(E)\sim-\displaystyle\frac c\pi v_0}{fapp}
Unfortunately, it turns out that the values of the energy for which
the above equation is satisfied are not physical, as it will be shown
below.

Other useful information on the influence of repulsive forces on the
winding angle can be obtained studying the form of the function
$G_1(E;\mathbf u,\mathbf v)$ of Eq.~\ceq{nontrainvpro}. We remember in
fact that all the effects of the repulsive forces are concentrated in
this component of the propagator.
Supposing for example
that the value of $|\mathbf u|$ is very large, i.~e.:
\beq{
|\mathbf u|\gg\frac 1{\sqrt{2Ec}}
}{cslm}
we have the following approximate expression of $G_1(E;\mathbf
u,\mathbf v)$:
\beq{
G_1(E;\mathbf u,\mathbf v)=\frac c{\sqrt{2\pi}}\lambda(E)
(2Ec)^{1/4}e^{-\sqrt{2Ec}|\mathbf u|}K_0(\sqrt{2Ec}|\mathbf v|) 
}{appexplx}
A relation analogous to Eq.~\ceq{appexplx} may  be
written also for the variable $\mathbf v$.
In practice, Eq.~\ceq{appexplx} means that the repulsive interactions
do not play any particularly relevant role in polymer configurations
in which the ends of the trajectories at some point are very distant.
%
This is not a surprise. If the trajectories at some height $t$ are very
far from each other, they will have little or no chance to interact
together via the repulsive interactions of Eq.~\ceq{dirpot}, which are
of short range. Eq.~\ceq{appexplx}
gives the concrete law with which the contributions of the repulsive
forces are suppressed in configurations of this kind.
In particular,
if the distance between the trajectories  is much greater
than the
 {\it
  characteristic length scale} 
\beq{l_{rep}=1/\sqrt{2Ec}}{chalensca}
the influence of the repulsive forces
ceases to be relevant. 
Of course, even if at some
points the 
trajectories are very distant, polymers will always have a chance to get
near enough to be able to interact if they are sufficiently long.
As a consequence, we expect that the characteristic length $l_{rep}$ 
increases with the increasing of the lengths of the trajectories. It
is easy to check that this is exactly the case. 
To show that, let us consider the dependence of $l_{rep}$ on
the polymer length. 
One parameter which determines this length is the distance $L$
between 
the ends of the polymers along
the $t-$axis. Indeed, a trajectory connecting the two ends of a
polymer must be
very long if these ends are
located at very distant heights. In the energy representation, large
values of $L$ correspond to small values of $E$. 
For example, in the limit
$E=0$, which
corresponds to infinite polymer lengths, we have that
$l_{rep}=\infty$, confirming our intuitive expectations.
Another confirmation comes from Eq.~\ceq{randisell} below, where
a rough estimation of the behavior of $l_{rep}$ with respect to the
distance $L$ is given. The dependence on $L$ is not the whole
story. As a matter of fact,
during their random walk in the $t-$direction from $t=0$ to $t=L$,
polymers are also allowed to wander in the remaining two
directions. Loosely speaking, the variations in the length of the
trajectories associated to the fluctuations in these transverse
directions are taken into account by the parameter
$c$. Smaller values of $c$ correspond to longer trajectories and
vice-versa, see \cite{ferrari}. It is now easy to
realize from Eq.~\ceq{chalensca} that the 
characteristic length $l_{rep}$ increases when $c$ decreases as expected.
Taking into account all the above considerations, it is
possible to conclude that repulsive forces give relevant contributions
to the second moment only in the case of configurations of the system
in which 
the trajectories of the two polymers are not too far from each other.
As a matter of fact, in the propagators appearing in the amplitudes of
Eqs.~(\ref{amponek}--\ref{amptwofour})
all configurations in
which the distance between the 
trajectories at some height in the $t-$axis is bigger than a few
characteristic 
lengths $l_{rep}$ are exponentially suppressed according to
Eq.~\ceq{appexplx}. One may also add that this suppression becomes
milder in the case of long polymers, because we have seen that the
characteristic  
lengths $l_{rep}$ grows with the length of the polymers with a law
which has been given in Eq.~\ceq{randisell}.

In the rest of this Section we will analyze some interesting limiting
cases, in which repulsive interactions become particularly weak or
strong. To this purpose, it would be appealing to consider the quantity
$c^{-1}\lambda(E)$, where $\lambda(E)$ has been given in
Eq.~\ceq{lambe},
  as an energy dependent effective or running coupling constant 
 of the repulsive interactions. This could be suggested by the
 expansions of Eqs.~(\ref{dexp}--\ref{expioe}) and by the fact
 that the quantity $c^{-1}\lambda(E)$ has the right dimension to be a
 coupling constant. 
Indeed, we will see that there are cases in which the strength of 
$\lambda(E)$ really determines the strength of the repulsive forces.
However, this is not true in general, as it should be because
 $\lambda(E)$ is just a parameter which has been factored out from the
expression of $G_1(E;\mathbf u,\mathbf v)$ and thus its meaning does
 not coincide with that of
a running coupling
constant.  
Keeping that in mind, we start to study the perturbative regime, 
in which
 $v_0$ is very small. 
In the
part of the propagator  in which there are the contributions of the
repulsive forces, i. e. the function
$G_1(E;\mathbf u,\mathbf v)$, $v_0$ is present only inside
$\lambda(E)$.
Expanding this quantity in powers of $v_0$, we obtain:
\beq{
\lambda(E)\sim \frac c\pi\left(-
v_0+\frac c\pi v_0^2K_0(\sqrt{2Ec}a)+\ldots
\right)
}
{pertapp}
We see that, at the leading order in $v_0$, $\lambda(E)$ is
 proportional to $v_0$ and thus, as it could have been expected,
$G_1(E;\mathbf u,\mathbf v)$ may be treated
 as a small perturbation with respect to the free
 propagator $G_0(E,\mathbf u,\mathbf v)$.
Let's now go back to Eq.~\ceq{fapp}. In that equation
it turns out that $\lambda(E)$ has the same
behavior as in the perturbative regime, even if Eq~\ceq{fapp} has been
derived in the hypothesis that $2Eca^2\ge 10$, but without supposing
that $v_0$ is small.
Before dwelling upon the physical meaning of this coincidence, let's
see what is the significance of the condition $2Eca^2\ge 10$.
To this purpose, we make the following approximations:
\beq{
L\sim
 E^{-1}\qquad\qquad\frac 1c\sim a
}
{draapp}
As mentioned before, it is quite reasonable to assume that the length
$L$ is proportional to the inverse of the energy $E$, while the second
approximation implies that polymers are very flexible. For example, in
polyethylene the Kuhn length$\sim 1/c$ is of the order of
molecular sizes. Exploiting Eq.~\ceq{draapp}, it turns out that the
condition $2Eca^2\ge 10$ is equivalent to the condition $L\le \frac
a5$. This would mean that our system is squeezed in a
volume whose height $L$ is smaller than the size of a monomer. 
Clearly, this situation is not very physical.

Since we have been able to compute the exact form of the propagator
$G(E;\mathbf u,\mathbf v)$, it is not difficult to study also the strong
coupling limit $v_0\longrightarrow\infty$.
As in the perturbative case, the only affected part 
of the propagator \ceq{gffin}
is the factor $\lambda(E)$
appearing in $G_1(E;\mathbf u,\mathbf v)$.
After a trivial calculation, one finds that, in the strong coupling
limit, the form of $G_1(E;\mathbf u,\mathbf v)$ is given by:
\beq{
G_1(E;\mathbf u,\mathbf v)\sim\left(K_0(\sqrt{2Ec}a)
\right)^{-1}
K_0(\sqrt{2Ec}|\mathbf u|)
K_0(\sqrt{2Ec}|\mathbf v|)
}{scnontrainvpro} 
Assuming that polymers are very long,
let us study the left hand side of the above
equation. This is a ratio of
modified Bessel functions of the second 
kind.
Since $a$ is a very small quantity and these functions have a
logarithmic singularity if their argument is small, see Eq.~\ceq{fdsf},
it is licit to suppose that
\beq{K_0(\sqrt{2Ec}a)>K_0(\sqrt{2Ec}|\mathbf u|)
K_0(\sqrt{2Ec}|\mathbf v|)
}{dendomnum}
unless $|\mathbf u|\sim a$ and/or $|\mathbf v|\sim a$. 
On the other side, we know from Eq.~\ceq{cslm} that, if $|\mathbf u|,
|\mathbf v|\gg\frac 1{\sqrt{2Ec}}$, the product of modified Bessel
functions $K_0(\sqrt{2Ec}|\mathbf u|)
K_0(\sqrt{2Ec}|\mathbf v|)$ decays exponentially. 
In other words, in the left hand side of Eq.~\ceq{scnontrainvpro} the
denominator  will dominate over the numerator whenever the distance
between the polymer trajectories is not of the order of a few
molecular sizes. Thus, if $v_0$ is large, the major
contributions to winding angle coming from 
the repulsive interactions occur when the trajectories are very
near to each other. This could be expected from the fact that, in the
strong coupling limit, one recovers the excluded volume interactions.

Finally, let us study the domain of the parameters $E$ and $c$ in which
the condition 
\beq{
2Eca^2\ll1}{condd}
is verified. 
We will see that this domain is particularly interesting, because if
condition \ceq{condd} is verified, the corrections
of the repulsive interactions to the entropy dominated behavior of
ideal polymers become relevant.
It has been already shown that under the assumption made in
Eq.~\ceq{condd}, the parameter 
$\lambda(E)$ is approximated as in Eq.~\ceq{logcorr}. Even if it is
not strictly necessary, 
we suppose here that $v_0$ has some finite
value, while polymers are so long that the following inequality is
satisfied:
\beq{
v_0^{-1}\ll-\frac c\pi\log\left(
\sqrt{2Ec}a
\right)
}
{fjfsfd}
This further assumption is to eliminate the dependence on $v_0$, which
could introduce confusion in the following discussion due
to possible interferences of condition \ceq{condd}  with those of the
perturbative and 
strong coupling regimes.
In the $L-$space, Eq.~\ceq{fjfsfd}
corresponds to the inequality
$e^{{2\pi}\slash{cv_0}}\ll\frac 
 L{2a}$.
Now $G_1(E;\mathbf u,\mathbf v)$ may be approximated as
 follows: 
\beq{
G_1(E;\mathbf u,\mathbf v)\sim \frac c{\pi\log(\sqrt{2Ec}a)}
K_0(\sqrt{2Ec}|\mathbf u|)K_0(\sqrt{2Ec}|\mathbf v|)
}
{domhigene}
As promised, the above equation does not contain the parameter $v_0$.
We see from the left hand side of Eq.~\ceq{domhigene} that
the function $G_1(E;\mathbf u,\mathbf v)$ is logarithmically
 suppressed, due to the presence of $\log(\sqrt{2Ec}a)$ in the
 denominator.
This suppression effect is counterbalanced only at short distances
by the two modified Bessel
 functions of the second kind appearing
in the numerator, which diverge
logarithmically whenever $\sqrt{2Ec}|\mathbf u|=0$
and/or $\sqrt{2Ec}|\mathbf v|=0$. The total result of these
opposite effects in the expression of the averaged second moment will
be presented in Section~\ref{sec:six} after performing the explicit
computation of the amplitudes of Eqs.~(\ref{amponek}--\ref{amptwofour}).

To conclude this Section, let us give some concrete values of the
involved parameters.
First of all, let us
 estimate  the values of $L$,  for which the two polymer system
is in the regime \ceq{condd}.
Using the approximations made in Eq.~\ceq{draapp},  we may
conclude that, if the relation \ceq{condd} is satisfied,
  the length $L$  needs such that $L\gg 2a$, i.~e.
$L$ is at least of the order of
 hundred molecular lengths or more:
$L> 100a $. Moreover,
it is possible to give a rough estimation of the maximum distance of
the end points, after which the two polymers are too far from each
other to allow a relevant contribution to the winding angle
due to repulsive interactions. Using Eq.~\ceq{cslm}, in fact, it turns out  
that
the repulsive interactions
are relevant only in 
 the range of distances: 
\beq{
|\mathbf u|\ll\sqrt{\frac {La}2}\sim l_{rep}}{randisell}
Finally, the situation opposite to condition \ceq{condd} is not
 realistic, because it
leads to the constraint $L\ll
 2a$. This
 would corresponds to the case of a polymer
 which is shorter than the size of the molecules composing it.

\section{Calculation of the averaged second moment}\label{sec:six}

At this point we are ready to compute the quantities $N(E)$ and $D(E)$
of Eqs.~\ceq{ennee} and \ceq{deee}.
We start with $D(E)$. Using Eqs.~\ceq{deeeexp}, \ceq{dexp} and the splitting
\ceq{propsplit} of the
propagator, one has at the zeroth order in $\lambda(E)$:
\beq{
D^{(0)}(E)=\int d^2 r_0\int d^2 r_1 G_0(E;\br_1,\br_0)
=\int d^2 r_0\int d^2 r_1 \frac c\pi K_0(\sqrt{2Ec}|\br_1-\br_0|)
}
{zerordprdz}
After a shift of variables, the above equation gives:
\beq{D^{(0)}(E)=S\int d^2r_1\frac c\pi K_0(\sqrt{2Ec}|\br_1|)
}
{deezero}
where $S=\int d^2r_0$ is the total surface of the system in the two
dimensional
space, which is transverse to the $t-$axis.
Using the identity
\beq{
\int d^2 r_1\frac c\pi K_0(\sqrt{2Ec}|\br_1|)=\frac 1E
}
{intbas}
one finds:
\beq{
D^{(0)}(E)=S/E
}
{deezerointe}
This expression of $D^{(0)}(E)$ has the following interpretation: We are
performing here an 
average of the second moment
with respect to all possible
initial and final positions of the endpoints of the polymers and
$D(E)$  {\it counts} the number of these configurations. The component
 $D^{(0)}(E)$ of $D(E)$ depends only on the free propagator $G_0(E;\bx,\by)$,
which is translational invariant in the sense discussed after
Eq.~\ceq{gzspec}. This invariance explains
why the number of configurations grows proportionally to the surface $S$.
The reason is that, for each configuration of the polymers, one can
obtain other equivalently probable configurations by the symmetric 
 translation of their ends on the
surface $S$  at the initial and final instants.
Let us  now apply to $D^{(0)}(E)$ an inverse Laplace transform, in
order to go back to the $L-$space. After a simple calculation we obtain:
\beq{
D^{(0)}(L)=S
}
{dzerol}
i. e. $D^{(0)}(L)$ does not depend on $L$.

The next and last contribution to $D(E)$ is given by:
\beq{
D^{(1)}(E)=\int d^2 r_0 d^2 r_1 
G_1(E;\br_1,\br_0)=\int d^2 r_0 d^2 r_1 \frac c\pi
\lambda(E)K_0(\sqrt{2Ec}|\br_1|) K_0(\sqrt{2Ec}|\br_0|) 
}
{donestart}
Exploiting Eq.~\ceq{intbas} to integrate out the variables $\mathbf r_0$
and
$\mathbf r_1$, we get:
\beq{
D^{(1)}(E)=\frac \pi c\lambda(E)E^{-2}
}
{contdeeone}
We remark  that the above contribution to $D(E)$ vanishes in
the limit $v_0=0$. This could be expected due to the fact that
$D^{(1)}(E)$ collects all contributions coming from the repulsive
interactions. These interactions break explicitly
the translational invariance of the free part of the action and, as a
consequence, $D^{(1)}(E)$ is no longer proportional to the surface $S$ as
$D^{(0)}(E)$. Unfortunately, it is not easy to compute the inverse Laplace
transform of $D^{(1)}(E)$ 
without making some approximation. To this purpose, we assume that the
repulsive interactions are weak, i.~e. $v_0\ll 1$, and that the value
of $L$ is large. In this case, since  we are in the domain of small
$E$'s, it is possible to expand $D^{(1)}(E)$ up to the second order in
$v_0$ as follows:
\beq{
D^{(1)}(E)\sim\frac \pi c \left(
\frac c\pi E^{-2} v_0-\left(\frac c \pi v_0\right)^2E^{-2}\log(\sqrt{2Ec}a)
\right)
}
{deeoneexp}
In order to obtain the above equation we have used both Eqs.~\ceq{fdsf} and
\ceq{pertapp}.
The inverse Laplace transform of Eq.~\ceq{deeoneexp}
gives:
\beq{
D^{(1)}(L)\sim\left[ v_0-\frac c\pi v_0^2\left(
\log(\sqrt{2c}a)+\frac{C-1}2
\right)
\right]L+\frac c{2\pi} v_0^2L\log L
}{deeonel}
where $C\sim 0.577215664$ is the Euler constant.

Putting Eqs.~\ceq{deezerointe} and \ceq{contdeeone} together, we
obtain:
\beq{
D(E)=D^{(0)}(E)+D^{(1)}(E)=SE^{-1}+\frac\pi c\lambda(E)E^{-2}
}
{deeefin}
This is an exact result. An approximated expression of $D(L)$ can be
derived instead from Eqs.~\ceq{dzerol} and \ceq{deeonel}.

Now we turn to the derivation of $N(E)$. We start by computing order
by order in $\lambda(E)$ the
contributions to the quantities $K(E)$ and $I_\omega(E)$ of
Eqs.~\ceq{expke}
and \ceq{expioe} respectively.
At the zeroth order we have for $K(E)$:
\beq{
K^{(0)}(E)=\frac c{2\pi^2}\int d^2x\bA^2(\bx)
\int d^2 r_1
K_0(\sqrt{2Ec}|\br_1-\bx|)\int d^2r_0K_0(\sqrt{2Ec}|\br_0-\bx|)
}
{kzermain}
After performing
an easy integrations over the coordinates $\br_0,\br_1$, one
obtains:
\beq{
K^{(0)}(E)=\frac 1{2c}E^{-2}\int d^2 x\bA^2(\bx)
}
{kzerot}
The remaining integral with respect to the $\bx$ coordinate 
is both ultraviolet and infrared divergent and
needs to be regulated. We have already seen that the
 singularities in the ultraviolet domain may
consistently be eliminated with the
 introduction  of the
 small distance cut-off $a$. A large distance cut-off is 
 instead motivated by the fact that the size of a real system is
necessarily  finite. Implicitly, we have already 
 used this kind of infrared regularization in Eq.~\ceq{deezero}, where
 we have assumed that the total surface $S$ of the system in the
 directions which are transverse to the $t-$axis is finite.
Supposing that the shape of $S$ is approximately a disk of radius $R$, so that
$S\sim\pi R^2$,
we may write:
\beq{
\int d^2 x\bA^2(\bx)=\frac 1{2\pi}\int_a^R\frac {d\rho}\rho
}
{polinfultint}
Substituting Eq.~\ceq{polinfultint} in Eq.~\ceq{kzerot},
one obtains the following expression of
$K^{(0)}(E)$:
\beq{
K^{(0)}(E)=\frac 1{8\pi c}E^{-2}\log\left(
\frac{S}{a^2\pi}
\right)
}
{kzerfin}
 The inverse Laplace transform of $K^{(0)}(E)$ gives:
\beq{
K^{(0)}(L)=\frac{L}{8\pi c}\log\left(
\frac{S}{a^2\pi}
\right)
}{kzerfinl}

We have now to compute the quantities $I_\omega^{(0)}(E)$, with
$\omega=1,\ldots,4$. The expressions of the $I^{(0)}_\omega(E)$'s may be
obtained from Eqs.~\ceq{ionee} and
(\ref{amptwoone}--\ref{amptwofour}), by substituting everywhere the
propagator $G(E;\bx,\by)$ with its free version $G_0(E;\bx,\by)$.
It is easy to show that:
\beq{
I_\omega^{(0)}(E)=0\qquad \qquad \mbox{for $\omega=1,\ldots,4$}
}
{vanio}
This vanishing, which is actually a
double vanishing, is due  to the fact that
each of the
$I^{(0)}_\omega(E)$'s contains an integral of
a total
divergence together with an integral which is zero for symmetry reasons. 
For some values of $\omega$, like for instance when $\omega=3$, to
isolate such 
integrals it is necessary to perform some integrations by parts. This
is allowed because the $I_\omega^{(0)}(E)$'s are not affected by
divergences,  contrarily to $K(E)$.

As an example, we work out explicitly the case of
$I^{(0)}_1(E)$. The first vanishing integral is the following:
\beq{
\int d^2 r_0\nabla^j_{\sby}
G_0(E;\br_0,\by)=
\frac c\pi\int d^2 r_0\nabla^j_{\sby}K_0(\sqrt{2Ec}|\br_0-\by|)
}
{inttotdivone}
This is of course zero due to symmetry reasons.
The second vanishing integral in $I^{(0)}_1(E)$
is of the form:
\beq{
I=\int d^2x\int d^2r_1 A_i(\bx)G_0(E;\br_1,\bx)\nabla^i_{\sbx}G_0(E;\by,\bx)
}{fff}
After performing the integration over $\br_1$ with the help of a shift
of variables and of Eq.~\ceq{intbas}, we have, apart from a
proportionality factor:
\beq{
I\propto\int d^2xA_i(\bx)\nabla^i_{\sbx}G_0(E;\by,\bx)
}
{deeint}
Since $A_i(\bx)$ is a divergenceless vector potential,
i.~e. $\nabla^i_{\sbx}A_i(\bx)=0$,
$I$ can be rewritten as the integral of a total divergence:
\beq{I=\frac c\pi\int d^2x\nabla^i_{\sbx}\left(
A_i(\bx)K_0(\sqrt{2Ec}|\by -\bx|)
\right)
}{inttotdivtwo}
Clearly, the left hand side of the above equation is zero. 
This fact can be also checked  passing to the Fourier
representation. Exploiting Eq.~\ceq{ftsmbf} 
and the formula
\beq{
A^i(\bx)=\frac 1{(2\pi)^2i}\int d^2\bp \epsilon^{ij}\frac{p_j}{\bp^2}
e^{i\sbp\cdot\sbx}
}{vecpotfourep}
in Eq.~\ceq{deeint}, one obtains:
\beq{
I=-\frac 1{(2\pi)^2}\int
d^2\bp\frac{\epsilon^{ij}p_ip_j}
{(\bp^2+2Ec)\bp^2}
}
{deeintiszero}
Thus $I=0$ because $\epsilon^{ij}p_ip_j=0$.
In an analogous way one shows that also $I^{(0)}_2, I^{(0)}_3$ and
$I^{(0)}_4$ are identically equal to zero. 

We are now ready to compute the contributions to $N(E)$, which are
linear in $\lambda(E)$. First of all, we treat the term $K^{(1)}(E)$,
which is given by:
\beq{
K^{(1)}(E)=\frac 1{2c}\int d^2x\int d^2 r_0\int d^2 r_1 \bA^2(\bx)
\left[
G_1(E;\br_1,\bx)G_0(E;\bx,\br_0)+
G_0(E;\br_1,\bx)G_1(E;\bx,\br_0)
\right]
}
{konee}
The
 integrations over $\br_0$ and $\br_1$ may be easily performed using
Eq.~\ceq{intbas} and give as a result a factor which is proportional
to
$E^{-2}$.
After that, only
the following integral in
$\bx$ remains to be done:
\beq{
\int d^2x\bA^2(\bx)K_0(\sqrt{2Ec}|\bx|)\equiv
\frac 1{(2\pi)^2}\int_{|\sbx|\ge a}d^2x\frac 1{|\bx|^2}K_0(\sqrt{2Ec}|\bx|)
}
{fserf}
Here the ultraviolet divergence, which is present  
in the left hand side, has been regulated in the usual way with the
introduction of the short distances cut-off $a$. Infrared divergences
are absent.
Going to polar coordinates, the right hand side of the
 above equation becomes:
\beq{
\frac 1{(2\pi)^2}\int_{|\sbx|\ge a}d^2x\frac 1{|\bx|^2}K_0(\sqrt{2Ec}|\bx|)=
\frac1{2\pi}\int_a^{+\infty} d\rho\frac{K_0(\sqrt{2Ec}\rho)}{\rho}
}
{fserfpolcoo}
Putting everything together, one arrives at the final result:
\beq{
K^{(1)}(E)=\frac 1{2\pi c}E^{-2}\lambda(E)\int_a^{+\infty}
d\rho\frac{
K_0(\sqrt{2Ec}\rho)
}{\rho}
}
{konefin}
If  the quantity $\sqrt{2Ec}a$ is small,
it is possible to 
derive
the following asymptotic expression of $K^{(1)}(E)$:
\beq{
K^{(1)}(E)\sim\frac 1{4\pi c}E^{-2}\lambda(E)
\log^2(\sqrt{2Ec}a)
}{koneeer}
To go from Eq.~\ceq{konefin} to Eq.~\ceq{koneeer}, we have used
the asymptotic formula:
\beq{
\int_a^{+\infty}
d\rho\frac{
K_0(\sqrt{2Ec}\rho)
}{\rho}\sim\frac 12\log^2(\sqrt{2Ec}a)
}
{leaordaasyexp}
which is valid for small values of $\sqrt{2Ec}a$.
We see from Eqs.~\ceq{konefin} and \ceq{leaordaasyexp}
that the presence of ultraviolet
divergences, together with the needed regularization, has modified the
naive form of $K^{(1)}(E)$ as a function of the pseudo-energy $E$
given in Eq.~\ceq{behaone}. The modification consists in the
appearance of the factor
$\int_a^{+\infty}\frac{d\rho}{\rho}K_0(\sqrt{2Ec}\rho)$, which
exhibits a square logarithmic singularity in the limit $\sqrt{2Ec}a=0$.

The inverse Laplace transformed of
$K^{(1)}(E)$ can be 
derived only 
making some approximation. As in the case of $D^{(1)}(E)$, we
will work
in the double limit, in which $v_0$ is very small and $L$ is very
large. After a few calculations we obtain:
\begin{eqnarray}
K^{(1)}(L)
&\sim&\frac {v_0}{4\pi^2}
\left\{
\frac 14 \int_0^Lds\left[
\log(L-s)+C
\right]\left(
\log s+C
\right)+\right.\nonumber\\&&\left.
\frac14L\log^2(2ca^2)+\frac 12 \log(2ca^2)
\left[
(C-1)L-L\log L
\right]
\right\}
\label{ddd}
\end{eqnarray}

At this point we have to compute the expressions of the $I_\omega^{(1)}(E)$'s,
$\omega=1,\ldots, 4$.
It is possible to show that these contributions  vanish
identically, i.~e.:
\beq{
I^{(1)}_\omega(E)=0\qquad\qquad\mbox{for $\omega=1,\ldots,4$}
}{vaniome}
The motivations of this vanishing are similar to the motivations for
which there are no contributions at the zeroth order: All terms which
appear in the
quantities $I_\omega^{(1)}(E)$ contain
at least one integral of a total 
divergence or one integral, which is zero for dimensional reasons.
As in the case of the  $I_\omega^{(0)}(E)$'s, there are some values of
$\omega$ for which it is necessary to perform an integration by parts
in order to isolate these vanishing integrals. Once again, this is
allowed because the  $I_\omega^{(1)}(E)$'s do not contain divergences.

At the next order in $\lambda(E)$, we have the last contribution to
$K(E)$:
\beq{
K^{(2)}(E)=\frac 1{2c}\int d^2x\int d^2 r_1\int d^2 r_0
\bA^2(\bx)G_1(E;\br_1,\bx)G_1(E;\bx,\br_0)
}{lcke}
After performing the integrations in $\br_1$ and $\br_0$ with the help
of Eq.~\ceq{intbas}, Eq.~\ceq{lcke} becomes:
\beq{
K^{(2)}(E)=
\frac \pi c\lambda^2(E)E^{-2}\int d^2x
\bA^2(\bx)\left(K_0(\sqrt{2Ec}|\bx|)\right)^2
}{erewf}
The integral in $\bx$ is divergent and needs a
regularization. Going to polar coordinates, we obtain the result:
\beq{
K^{(2)}(E)\equiv
\frac \pi c\lambda^2(E)E^{-2}\int_a^{+\infty}\frac{d\rho}{\rho}
\left(K_0(\sqrt{2Ec}\rho)\right)^2
}{ktwofin}
Also in this case, we note that the presence of the regularization
modifies
the 
dependence of $K^{(2)}(E)$ on the pseudo-energy $E$
with respect to the naive formula of Eq.~\ceq{behaone}.
The correction consists in the factor
$\int_a^{+\infty}\frac{d\rho}{\rho}
\left(K_0(\sqrt{2Ec}\rho)\right)^2$.
In the limit
$\sqrt{2Ec}a=0$, this factor diverges as powers of $\log(\sqrt{2Ec}a)$.

To conclude the analysis of the contribution to $N(E)$ at the second
order in $\lambda(E)$, we show that the $I_\omega^{(2)}(E)$'s are
identically equal to
zero. As a matter of fact, it is easy to verify that for
$\omega=1,2,4$ each $I_\omega^{(2)}(E)$ contains terms of the kind:
\beq{
B(\bx)=A_i(\bx)\nabla^i_{\mathbf x}K_0(\sqrt{2Ec}|\bx|)
}{blmn}
These terms vanish identically because of the following
identity:
\beq{
\nabla_{\mathbf x}^iK_0(\sqrt{2Ec}|\bx|)=
\frac 1{\sqrt{2Ec}}\frac{x^i}{|\mathbf x|^2}\frac{\partial
K_0(\sqrt{2Ec}|\bx|) }{\partial|\mathbf x|}
}{auxfskj}
Substituting Eq.~\ceq{auxfskj} in Eq.~\ceq{blmn} and using the
explicit expression of the vector potential $A_i(\bx)$ of
Eq.~\ceq{vecpot}, we get: 
\beq{
B(\bx)=\frac 1{2\pi\sqrt{2Ec}}\frac{\epsilon_{ji}x^ix^j}{|\mathbf x|^4}
\frac{\partial
K_0(\sqrt{2Ec}|\bx|) }{\partial|\mathbf x|}
}
{dfsdf}
Clearly, the right hand side of the above equation is zero because
$\epsilon_{ji}x^ix^j=0 $.
If $\omega=3$, instead, the vanishing function $B(\bx)$ of
Eq.~\ceq{blmn} may be isolated in the expression of 
$I_3^{(2)}(E)=0$ only after
an integration by parts.

Finally, at the third order in $\lambda(E)$ we have
only the
quantities $I_\omega^{(3)}(E)$'s, since $K(E)$ has at most quadratic
powers of $\lambda(E)$. It is easy to
realize that:
\beq{
\sum_{\omega=1}^4I_\omega^{(3)}(E)=0
}
{threezero}
because the following relations hold \footnote{Let us stress the fact
  that each 
  of the $I_\omega^{(3)}(E)$'s is separately equal to zero, because these
  quantities contain terms of the kind given in 
  Eq.~\ceq{blmn}.}: 
\beq{
I_1^{(3)}(E)=-I_2^{(3)}(E)=I_3^{(3)}(E)=-I_4^{(3)}(E)
}
{dsfsfs}
As a consequence of Eq.~\ceq{threezero},  it is clear that there are
no contributions to $N(E)$ at this order.

Using
Eqs.~\ceq{kzerfin}, \ceq{konefin} and \ceq{ktwofin}, we arrive at the
 final 
result for $N(E)$:
\begin{eqnarray}
N(E)&=&
\frac 1{4\pi c}E^{-2}\log\left(
\frac S{a^2\pi}
\right)+\frac
1{\pi c}\lambda(E)E^{-2}\int_a^{+\infty}\frac{d\rho}{\rho}
K_0(\sqrt{2Ec}\rho)\nonumber\\
&+&\frac {2\pi} c\lambda^2(E)E^{-2}\int_a^{+\infty}\frac{d\rho}{\rho}
\left(K_0(\sqrt{2Ec}\rho)\right)^2\label{nefin}
\end{eqnarray}

We can now insert in the formula of the second moment of
 Eq.~\ceq{secmomaveerep} the functions $D(E)$ and
$N(E)$ given in Eqs.~\ceq{deeefin} and \ceq{nefin} respectively.
The outcome is:
\beq{
\langle m^2\rangle(E)=
\frac{
E^{-1}\left[
\frac 1{4\pi c}\log\left(
\frac S{a^2\pi}\right)
+
\lambda(E)
\frac {
\int_a^{+\infty}\frac{d\rho}{\rho}
K_0(\sqrt{2Ec}\rho)
}{\pi c}
+
\lambda^2(E)\frac {2\pi} c\int_a^{+\infty}\frac{d\rho}{\rho}
\left(K_0(\sqrt{2Ec}\rho)\right)^2
\right]
}{
S+\frac \pi c\lambda(E)E^{-1}
}
}
{mefin}
In the $L-$space, the already mentioned difficulties with the
computation of the inverse 
Laplace transform of $D(E)$ and $N(E)$ allow an analytical result only
in the double limit of weak coupling constant $v_0$ and of large
values of $L$. At the first order in $v_0$, the expression of
$\langle m^2\rangle$ reads as follows:
\beq{
\langle m^2\rangle=\frac{
\frac L{8\pi c}\log\left(
\frac S{a^2\pi}\right)+K^{(1)}(L)
}{S+v_0L}
}
{mlfin}
where $K^{(1)}(L)$ has been given in Eq.~\ceq{ddd}.

So far, we have considered the averaged second moment
of Eq.~\ceq{secmomerep}, corresponding to the case in which the polymer
ends are not fixed.
In the energy representation, we have seen that this
version of the second moment can be
exactly computed. To conclude this Section, we would like to  show
that it is 
possible to provide also an exact expression of the second moment
$\langle m^2\rangle_{\mathbf r_1,\mathbf r_0}$ in the $L-$ space and
with fixed polymer ends up to an inverse Laplace transform of the
propagator given in Eq.~\ceq{gffin}.
The starting point is the exact formula of the second moment $\langle
m^2\rangle_{\br_1,\br_0}(E)$ of Eq.~\ceq{feesm}.
All the ingredients of this formula are defined in
Eqs.~\ceq{secmomerep}, 
\ceq{nsdoqop}--\ceq{nsdtqop} and \ceq{ampzero}--\ceq{amptwofour}.
Looking at Eq.~\ceq{feesm}, it is clear that:
\beq{
N(E;\br_1,\br_0)=2K(\br_1,\br_0)-\sum_{\omega=1}^4I_\omega(\br_1,\br_0)
}
{nefull}
and
\beq{
D(E;\br_1,\br_0)=G(E;\br_1,\br_0)
}
{defull}
Let us note that the functions $I_\omega(\br_1,\br_0)$ are all equal
up to integrations by parts, which can shift the differential
operators $\nablab_{\sbx}$ and $\nablab_{\sby}$ in
Eqs.~\ceq{amponek}--\ceq{amptwofour}. 
This fact will be used  in order to simplify the expression of
the inverse Laplace transformed of 
$N(E;\br_1,\br_0)$ in the $L-$space.
To compute the inverse Laplace transforms of
both $N(E;\br_1,\br_0)$  
and $D(E;\br_1,\br_0)$,
we use the following property of the
inverse Laplace transform of the product of two functions
$f(E)$ and $g(E)$:
\beq{
{\cal L}^{-1}(f(E)g(E))=\int_0^Ldsf(L-s)g(s)
}
{convilp}
Applying Eq.~\ceq{convilp} to evaluate the inverse Laplace transforms
of $K(E)$ and of the $I_\omega(\br_1,\br_0)$ in
Eqs.~\ceq{nefull} and \ceq{defull}, we
obtain after some calculations:
\begin{eqnarray}
N(L;\br_1,\br_0)&=&\frac 2c\int d^2x \bA^2(\bx)\int_0^L ds
G(L-s;\br_1,\bx) G(s;\bx,\br_0)\nonumber\\
&-&\frac 2{c^2}\int d^2x\int d^2 y\int_0^L G(L-s;\bx,\br_1)\nonumber
\\
&\times& \int_0^s ds'\partial_x^i\partial_y^jG(s-s';\by,\bx)
G(s';\br_0,\by)A_i(\bx)A_j(\by)
\label{nlfull}\\
D(L;\br_1,\br_0)&=&G(L;\br_1,\br_0)\label{dlfull}
\end{eqnarray}
The second term in the right hand side of Eq.~\ceq{nlfull} is the
contribution given by the functions $I_\omega(\br_1,\br_0)$,
$\omega=1,\ldots,4$, while the first term comes from $K(\br_1,\br_0)$.
Remembering the definition 
\ceq{secmomnumden}
of the second moment in terms of 
$N(L;\br_1,\br_0)$ and $D(L;\br_1,\br_0)$, we get:
\begin{eqnarray}
\langle m^2\rangle_{\br_1,\br_0}&=&
[G(L;\br_1,\br_0)]^{-1}\left[\frac 2c
\int d^2x \bA^2(\bx)\int_0^L
ds G(L-s;\br_1,\bx)G(s;\bx,\br_0)\right.\nonumber\\
&&\!\!\!\!\!\!\!\!\!\!\!\!\!\!\!\!\!\!\!\!\!\!\!\!\!\!
\!\!\!\!\!\!\!\!\!\!\!\!\!\!\!\!\!\!\!\!\!
\left.-\frac{2}{c^2}
\int d^2x\int d^2 y\int_0^Lds G(L-s;\bx,\br_1)\int_0^s ds'
\partial_x^i\partial_y^jG(s-s';\by,\bx) G(s';\br_0,\by)
A_i(\bx)A_j(\by)\right]\label{formtl}
\end{eqnarray}
If we knew how to compute the propagator $G(L;\bx,\by)$ starting from
its Laplace transformed \ceq{gffin}, we could evaluate explicitly the
expression of the second moment in the $L-$space.
Unfortunately, it is too complicated to perform
the inverse  Laplace transform of the
propagator $G(E;\bx,\by)$. Due to this technical difficulty,
Eq.~\ceq{formtl} is only formal. Progress can be made however in the limit
$v_0=0$, in which the propagator is given by the Green function
$G_0(E;\bx,\by)$ of Eq.~\ceq{gzspec}. This will be done in the next Section.


\section{The case of ideal polymers}\label{sec:seven}
In order to allow the comparison with previous results, this
Section is dedicated to the case of ideal chains in which $v_0=0$.
First of all, we discuss the
 formula of the averaged second moment
derived in the previous Section, Eq.~\ceq{mlfin}.
In the limit $v_0=0$,
Eq.~\ceq{mlfin} becomes:
\beq{
\langle m^2\rangle_0=
\frac L{8\pi cS}\log\left(
\frac S{a^2\pi}\right)
}
{mlzerofin}
The presence of a geometrical factor
like the surface $S$ of the system in the expression
of $\langle m^2\rangle_0$ 
has been already related to the
translational symmetry of Eqs.~\ceq{strainvzer} and \ceq{strainvone}.
Assuming that this surface has approximately the shape of a disc of
radius $R$, we
can put $S=\pi R^2$ as in Eq.~\ceq{polinfultint}.
Eq.~\ceq{mlzerofin} predicts that the average degree
of entanglement scales as follows with respect to the distance $R$:
\beq{
\langle m^2\rangle_0\propto
\frac{\log R}{R^2}}
{scalerisR}
The meaning of Eq.~\ceq{scalerisR} is the following. We remember that
the averaged second moment
$\langle m^2\rangle_0$ describes  the entanglement of two closed
polymers whose ends on the surfaces at $t=0$ and
$t=L$ are not fixed. 
In this way, the polymers are allowed to move freely and it is natural
to suppose that, the bigger will be the volume $SL$ in which the polymers
fluctuate, the bigger will be also the average distance between them.
Thus, if the surface $S$ increases its area, the probability of
entanglement must decrease. 
The exact law of this
decreasing is given by Eq.~\ceq{scalerisR}. 

On the other side, 
one would expect that
the probability of
getting
entangled is higher for
 long polymers 
than for short
polymers. 
Eq.~\ceq{mlzerofin} gives a result which is in
agreement with the above expectation, because  the second moment $\langle
m^2\rangle_0$ 
 scales as follows with respect to the parameters $L$ and $c$, which
 determine the polymer length:
\beq{
\langle m^2\rangle_0\propto
\frac{L}{c}}
{scalereslc} 
In particular, one can show  that the total length
of a polymer increases proportionally to $L$ and it is inversely
proportional to the square root of $c$ \cite{ferrari}.
Accordingly, we see from Eq.~\ceq{scalereslc} that  $\langle
m^2\rangle_0$ increases proportionally to $L$ and inversely
proportional to $c$.

At this point we wish to study the second moment 
$\langle
m^2\rangle_{0,\br_1,\br_0}$ 
of polymers with
fixed endpoints. 
The subscript $0$ has been added to the symbol of the second moment
to remember that we are working in the
limit $v_0=0$. 
Since we are dealing with ideal polymers, we have to substitute
everywhere in
Eq.~\ceq{formtl} the
full propagator $G(L;\bx,\by)$ with the free one. The result of this
operation is:
\begin{eqnarray}
\langle m^2\rangle_{0,\br_1,\br_0}&=&
[G_0(L;\br_1,\br_0)]^{-1}\left[\frac 2c
\int d^2x \bA^2(\bx)\int_0^L
ds G_0(L-s;\br_1,\bx)G_0(s;\bx,\br_0)-\right.\nonumber\\
&&\!\!\!\!\!\!\!\!\!\!\!\!\!\!\!\!\!\!\!\!\!\!\!\!\!\!
\!\!\!\!\!\!\!\!\!\!\!\!\!\!\!\!\!\!\!\!\!\!\!\!\!
\left.\frac{2}{c^2}
\int d^2x\int d^2 y\int_0^Lds G_0(L-s;\bx,\br_1)\int_0^s ds'
\partial_x^i\partial_y^jG_0(s-s';\by,\bx) G_0(s';\br_0,\by)
A_i(\bx)A_j(\by)\right]\label{mlfullvez}
\end{eqnarray}
We notice that, as it could be expected, Eq.~\ceq{mlfullvez} coincides
with the expression  obtained in \cite{tanaka} for the second moment
of one polymer winding up around an infinitely long straight wire
lying along the $z-$axis. 
Luckily, the propagator $G_0(L;\br_1,\br_0)$ can be explicitly constructed
 upon computing the
inverse Laplace transform  of the propagator $G_0(E;\br_1;\br_0)$
of Eq.~\ceq{gzspec}:
\beq{
G_0(L;\bx,\by)
=\frac c{2\pi L} e^{\frac c{2L}|\sbx-\sby|^2}
}
{freelprop}

It is easy to check that the second term in
the right hand side of Eq.~\ceq{mlfullvez}, which is associated with
the contributions coming from the $I_\omega(\br_1,\br_0)$'s, does not
grow with 
increasing values of $L$. As a matter of fact, after a
rescaling of variables, the numerator of this term gives:
\begin{eqnarray}
&&\frac{2}{c^2}
\int d^2x
\int d^2 y
\int_0^Lds G_0(L-s;\bx,\br_1)\int_0^s ds'
\partial_x^i\partial_y^jG_0(s-s';\by,\bx) G_0(s';\br_0,\by)
A_i(\bx)A_j(\by)\nonumber=\\
&&\frac c{4\pi^3 L}\int d^2x'\int d^2y'\int_0^1 dt\frac 1{1-t}
e^{-\frac c{2(1-t)}\left|\sbx'-\frac{\mathbf r_1}L\right|^2}
\int_0^tdt'\frac
1{t-t'}\left[
\frac{\partial^2}{\partial_{x_i'}\partial_{y_j'}}
e^{-\frac c{2(t-t')}\left|\mathbf x'-\mathbf y'\right|^2}\right]\nonumber\\
&&\frac 1{t'}
e^{-\frac c{2(t')}\left|\sby'-\frac{\mathbf r_0}L\right|^2}
\label{mlsmfullvez}
\end{eqnarray}
In the limit $L\longrightarrow\infty$, the quantity in the right hand
side of the above equation scales as $AL^{-1}$, where $A$ is a
constant. 
Moreover,
the propagator \ceq{freelprop}, which is in the denominator,
 scales as $L^{-1}$.
Thus, the ratio between the right hand side of Eq.~\ceq{mlsmfullvez} and
the propagator \ceq{freelprop} does not depend on $L$. This completes
the proof of our statement.

As a consequence of this statement, as far as the scaling of $\langle
m^2\rangle_{0,\br_1,\br_0}$ for large values of $L$ is concerned, it is
possible to make the following approximation:
\beq{
\langle m^2\rangle_{0,\br_1,\br_0}\sim
\frac 2c[G_0(L;\br_1,\br_0)]^{-1}
\int d^2x \bA^2(\bx)\int_0^L
ds G_0(L-s;\br_1,\bx)G_0(s;\bx,\br_0)
}
{star}
Unfortunately, despite of the fact that we are treating ideal
polymers,
the
integral in $d^2x$ appearing in the above equation is still
complicated and
requires some approximation to be evaluated analytically.
We will apply to this purpose the strategy used in Ref.~\cite{tanaka}
to compute the 
second moment of three dimensional polymers, adapting it to our
two-dimensional case. First of all, let us note that the integral in
\ceq{star} is ultraviolet divergent. However, the infrared
divergences which appeared in the energy representation are
absent. This is due to 
the  behavior of the propagator $G_0(L;\bx,\by)$, which is much milder
at infinity
than the behavior of the Green function
$G_0(E;\bx,\by)$.
To regulate the singularities at small distances, we proceed as usual
by introducing the cut-off $a$. After a rescaling of all variables
similar to that
of Eq.~\ceq{mlsmfullvez}, we get:
\begin{eqnarray}
&&\langle m^2\rangle_{0,\br_1,\br_0}\sim
\frac 2c[G_0(L;\br_1,\br_0)]^{-1}\times
\nonumber\\
&&
\!\!\!\!\!\!\!\!\!\!\!\!\!\!\!\!\!\!\!\!\!\!\!\!\!
\int_{|x'|\ge\frac{a\sqrt c}{\sqrt L}} \frac{d^2x'}L
\frac 1{\mathbf x'^2} 
\int_0^1
\frac{ds'}{s'(1-s')} \left(
\frac c{2\pi}
\right)^2
e^{-\frac 1{2(1-s')}\left|\mathbf x'-\mathbf r_1\sqrt{\frac
    cL}\right|^2}
e^{-\frac 1{2s'}\left|\mathbf x'-\mathbf r_0\sqrt{\frac cL}\right|^2}
\label{starst}
\end{eqnarray}
To go further, following \cite{tanaka}, we assume that the relevant
contribution to the integral in $d^2x'$ comes from a narrow region
around the singularity in $x'=0$. Thus,
we may put
\begin{eqnarray}
&&\int_{|x'|\ge\frac{a\sqrt c}{\sqrt L}} \frac{d^2x'}{\mathbf x'^2}
e^{-\frac 1{2(1-s')}\left|\mathbf x'-\mathbf r_1\sqrt{\frac
    cL}\right|^2}
e^{-\frac 1{2(1-s')}\left|\mathbf x'-\mathbf r_0\sqrt{\frac
    cL}\right|^2}\nonumber\\
&\sim&
2\pi\log\left(\sqrt{\frac
  Lc}a\right)
e^{-\frac 1{2(1-s')}\left|\mathbf r_1\sqrt{\frac
    cL}\right|^2}
e^{-\frac 1{2(1-s')}\left|\mathbf r_0\sqrt{\frac
    cL}\right|^2}\label{inteval}
\end{eqnarray}
After making the above crude approximation, we obtain:
\begin{eqnarray}
&&\langle m^2\rangle_{0,\br_1,\br_0}\sim
\frac c{\pi L}[G_0(L;\br_1,\br_0)]^{-1}\log\left(\sqrt{\frac
  Lc}a\right)
\int_0^1ds'\left[
\frac 1{1-s'}
e^{-\frac 1{2(1-s')}\mathbf r_1^2\frac
    cL}
e^{-\frac 1{2s'}\mathbf r_0^2\frac
    cL}+
\right.\nonumber\\
&&\left.
\frac 1{s'}
e^{-\frac 1{2(1-s')}\mathbf r_1^2\frac
    cL}
e^{-\frac 1{2s'}\mathbf r_0^2\frac
    cL}\right]
\label{starstar}
\end{eqnarray}
In deriving the above equation we have used the 
%
%
 simple relation $\frac 1{s'(1-s')}=\frac 1{(1-s')}+\frac 1{s'}$.
Let us now study the integral
\beq{
\tilde I=\int_0^1\frac{ds'}{s'}
e^{-\frac 1{2(1-s')}\mathbf r_1^2\frac
    cL}
e^{-\frac 1{2s'}\mathbf r_0^2\frac
    cL}
}{inttrian}
The other integral in $ds'$ appearing in \ceq{starstar} can be treated in the
same way after the change of variables $1-s'=t$.
It is not to allowed to take in the right hand side of
Eq.~\ceq{inttrian} the limit $L\longrightarrow\infty$ 
because in this
way the integral will not be convergent due to the singularity in
$s'=0$.
For this reason, we split the domain of integration as follows:
\begin{eqnarray}
\tilde I&=&\int_0^u\frac{ds'}{s'}
e^{-\frac 1{2(1-s')}\mathbf r_1^2\frac
    cL}
e^{-\frac 1{2s'}\mathbf r_0^2\frac
    cL}\nonumber\\
&+&\int_u^1\frac{ds'}{s'}
e^{-\frac 1{2(1-s')}\mathbf r_1^2\frac
    cL}
e^{-\frac 1{2s'}\mathbf r_0^2\frac
    cL}\label{intsplitted}
\end{eqnarray}
where $0< u < 1$.
Clearly, the second integral converges after performing the limit 
$L\longrightarrow\infty$ in the integrand and gives:
\beq{
\int_u^1\frac{ds'}{s'}=\log \frac 1u
}
{intconvlim}
The first integral instead diverges logarithmically with growing
values of $L$.
However, now it is possible to expand the exponential 
$e^{-\frac 1{2(1-s')}\mathbf r_1^2\frac
    cL}$ in powers of its argument, because the singularity in $s'=1$
lies outside the interval $[0,u]$.
Keeping only the leading order term with respect to $L$, we
get:
\beq{
\tilde I\sim -\mbox{Ei}\left(-\frac{u\mathbf r_0^2c}{2L}
\right)-\log u
}
{eifunc}
where Ei$(z)$ is the exponential-integral function. When $L$ is large,
this function may be approximated as follows: Ei$(z)\sim log(-z)$ and,
as a consequence:
\beq{
\tilde I\sim -\log\left(\frac{\mathbf r_0^2c}{2L}
\right)
}
{tildifin}
The second integral which we have left in Eq.~\ceq{starstar} gives the
same result.
Putting everything together in the expression of the second moment of
Eq.~\ceq{starstar}, we obtain the final result:
\beq{
\langle m^2\rangle_{0,\br_1,\br_0}\sim
-2\log\left(\sqrt{\frac
  Lc}a\right)
\log\left(\frac{\mathbf r_1^2\mathbf r_0^2 c^2}{4L^2}
\right)\sim 2\left(\log L\right)^2
}
{lllsecmomfreefin}
This is exactly the behavior of the second moment derived in
Ref.~\cite{tanaka}. 
\section{Conclusions}\label{sec:concl}
In this article we have studied the entanglement of two directed
polymers from a non-perturbative point of view. Our formulas
of the second moment,  
a quantity which measures the degree of entanglement
of the two polymers, 
take into account the repulsive forces acting on
the 
segments of the polymers and are exact.
The averaged second moment defined in Eq.~\ceq{aderr}, a version of
the second moment corresponding to the situation in which the end
points of the polymers are free to move, has been computed
in Eq.~\ceq{mefin}
as a function of the chemical potential $E$ conjugated to the distance
$L$ between the end points in the $t-$direction.
The case of free ends is
relevant in the treatment of nematic polymers and polymers in a
nematic solvent \cite{dirpoltheo}.
Let us note that also the expression of the
second moment without any averaging and in the $L$ space
can be computed. This has been done in
Eq.~\ceq{formtl}. However, this equation is explicit
only
up to the inverse Laplace
transform of the propagator \ceq{gffin}, which is too hard to be
obtained in closed form.


 Eq.~\ceq{mefin} shows that
the  
averaged second moment is of the
 form
 $\langle m^2\rangle(E) = E^{-1}f(E)$. The overall factor $E^{-1}$
 coincides with the scaling power law
of two ideal polymers. 
The correction $f(E)$ to this fundamental
behavior due to the repulsive interactions
is a complicated function of $E$, whose analysis would require
numerical methods. 
Nevertheless, it is possible to identify a
dominance of the repulsive interactions in the
domain of parameters in which the condition
$\sqrt{2Ec}a\sim 0$ is satisfied. This corresponds roughly speaking
to the situation in which polymers are very long.
In this region, the 
scaling laws with
respect to 
the energy $E$
of the numerator and denominator appearing in the right hand side of
Eq.~\ceq{mefin} are corrected by factors which are logarithmic powers
of $\log(\sqrt{2Ec}a)$, see for instance Eq.~\ceq{leaordaasyexp}.

One advantage of our approach is that it is easy to separate
within the expression
of the second moment 
the contributions of purely entropic origin which are typical of free
polymers from the contributions coming from the presence of the
$\delta-$function potential in the polymer action. This is essentially
due to the splitting \ceq{appexplx} of the
propagator $G(E;\mathbf u,\mathbf v)$ appearing in the
amplitudes (\ref{amponek}--\ref{amptwofour}). The component
$G_0(E;\mathbf u,\mathbf v)$  of the propagator coincides with the
propagator of ideal polymers, while the component
$G_1(E;\mathbf u,\mathbf v)$ 
takes into account the effects of the interactions.
Thanks to the splitting \ceq{appexplx}, it has been possible
 to study the way in which the
repulsive forces affect the 
average degree of entanglement of the two polymers.
This has been done in Section~\ref{sec:five}.
Our results are in agreement with the intuition. 
The  precise law with which the effects of the
repulsive forces on the entanglement decrease when the
distance between the trajectories increases is given by
Eq.~\ceq{appexplx}. 
 In Section~\ref{sec:five} it has been discussed also the 
strong coupling limit, which should be taken to recover
the limit of 
excluded volume interactions. 
In our exact approach, it is not difficult to consider the case in
which
the coupling constant $v_0$ is large. 
For instance, the component $G_1(E;\mathbf u,\mathbf v)$ of the
propagator, which is responsible of the effects due to the repulsive
interactions, has been given in the strong coupling limit in
Eq.~\ceq{scnontrainvpro}.
Studying the form of this component assuming that polymers are very
long, it has been argued 
that, at strong coupling,
the major
contributions to the winding angle coming from 
the repulsive interactions occur when the trajectories are very near
to each other.  
 Many other qualitative and quantitative characteristics of the
behavior of the two polymer system under consideration have been
presented in Section~\ref{sec:five}.

The case of ideal polymers, in which $v_0=0$, has been discussed
at the end of Section \ref{sec:seven} in order to  make comparison
with previous 
works. The scaling of the averaged
second moment for large values of $L$ obtained in
Eq.~\ceq{scalereslc} is in agreement with the results of
\cite{drossel}, if one takes into account the fact that, after the
averaging procedure of Eq.~\ceq{aderr} and the infrared regularization
of Eqs.~\ceq{deezero} and \ceq{polinfultint}. one is effectively treating a
system of  polymers confined in a cylinder of finite
volume $SL$.
In Section \ref{sec:seven} we have evaluated the second moment, always
of two ideal 
polymers, using the approach of Ref.~\cite{tanaka}.
The outcome of this calculation, namely the scaling behavior of
$\langle m^2\rangle_{0,\br_1,\br_0}$ at the leading order in $L$, is
reported in Eq.~\ceq{lllsecmomfreefin}.
This result is in agreement with the  square logarithmic behavior obtained
in \cite{tanaka}, but not with
the logarithmic behavior predicted in
\cite{drossel}. However, this discrepancy can be expected due to the
fact that, in 
Section \ref{sec:seven}, we have assumed, following
Ref.~\cite{tanaka}, that the most 
relevant contribution to the second moment
coming from the integral in Eq.~\ceq{starst}
is concentrated in a narrow region near the singularity in $x'=0$. 
This clashes with the assumptions of Ref. \cite{drossel}, 
in which instead it is argued that
the main increase in the winding
angle does not occur when the polymer trajectories are near, but rather when
they are far one from the other.
Finally, there is also an apparent
 discrepancy between the linear scaling with respect
to $L$ of the 
averaged second moment $\langle m^2\rangle_0$ 
and the square logarithmic scaling of the second moment $\langle
m^2\rangle_{0,\br_1,\br_0}$. This disagreement
is explained by the fact that, in the first case, the
ends of the polymers
are free to fluctuate, while in the second case they are fixed.
It is therefore licit to expect that two polymers with free ends are
more likely 
to entangle than two polymers whose ends are constrained.

Concluding, we would like to discuss possible further developments of
this work, together with some problems which are still left open.
First of all, 
 the number of
entangling polymers has been limited to two. To go
beyond this restriction, one should explore the possibility of
replacing the external vector potential $A_i(\bx)$ of Eq.~\ceq{vecpot}
with Chern--Simons fields. Abelian Chern-Simons field theories
have been already successfully applied
in order to impose topological constraints to the trajectories of
an arbitrary number of
closed polymer rings in \cite{fekllanpol}.
 We hope
to extend those results also to the case of directed polymers in a
forthcoming publication. Of course, if the polymer trajectories  are open,
the constraints among them are no longer of topological nature as in
\cite{fekllanpol}, so that the application of Chern-Simons field
theory to directed polymers should be considered with some care. 

We have also not made any attempt to
introduce in the treatment of polymer
entanglement more sophisticated
constraints than those which can be imposed with the help of the
winding angle. This is in effect still an unsolved problem, despite
the fact that two powerful and strategies have been
proposed 
for its solution  \cite{kleinertII,
  kholvil,nechaevII}.
In the first approach,
pioneered independently by Kleinert, Kholodenko and one of the authors
\cite{kleinert,kleinertII,kholvil}. the constraints are 
expressed via the Wilson loop amplitudes of non-abelian Chern-Simons field
theories. 
Some progresses toward a concrete realization of this program
in polymer physics have been made in 
Refs.~\cite{fernonsem,fernonsemII}. 
In the second approach, developed by Nechaev and coworkers, 
see \cite{nechaevII} and references therein.
polymer trajectories are mapped on a complex plane with punctures. The
link invariants necessary to impose the constraints are then
constructed using the properties of conformal maps.

Another possible development is the treatment of attractive
interactions, in which the strength $v_0$ in Eq.~\ceq{dirpot} takes
negative values. In this case, the $\delta-$function potential admits
a bound state \cite{jack} and the propagator of Eq.~\ceq{gffin}
develops a singularity, in which $\lambda(E)=\infty$, at the energy
corresponding to this bound state. It would be extremely interesting to
investigate how these facts affect the polymers' entanglement.
Another issue which deserves attention is that of hairpin turns. Hairpins
 are important in nematic solvents \cite{dirpoltheo} and can be
 included with the help of field theories \cite{cardy}.
We note also that in our formalism it is also possible to study
the entanglement of polymers in confined geometries. For example, values
of $E$ which are near to $a^{-1}$  ($E\le a^{-1}$) correspond roughly
speaking to the situation in which polymers fluctuate in a quasi
two-dimensional environment, in which the height in the $t-$direction
is of the order of a few molecular sizes.

Finally, an open problem, which has not be discussed here because 
 we were mainly
interested in the second moment.
is the derivation of
the full partition function 
${\cal G}_\lambda(E;\br_1,\br_0)$ 
of Eq.~\ceq{laplrep}. 
As anticipated in the Introduction, it is not an easy task to compute
${\cal G}_\lambda(E;\br_1,\br_0)$ 
 because the repulsive potential of Eq.~\ceq{dirpot}
is not central. 
We note however that the expression of
 ${\cal G}_\lambda(E;\br_1,\br_0)$ 
coincides with the Green function
of a spin $1/2$ Aharonov-Bohm problem in the imaginary time
formulation of quantum mechanics. This Green function has been already
derived in \cite{Park} using sophisticated techniques developed in
Refs.~\cite{hagen, jack}, which bypass all the difficulties of dealing
with a non-central potential. Thus, in principle, the expression of
 the partition 
function ${\cal G}_\lambda(E;\br_1,\br_0)$  is known.
Unfortunately, some
of the consistency conditions imposed on the parameters in Ref.~\cite{Park}
seem to be
incompatible with the requirements of our physical problem, as noted
in Section \ref{sec:three}.
For these reasons, the  computation of the full partition function
 ${\cal G}_\lambda(E;\br_1,\br_0)$ 
is still a problem which needs further investigations.
Luckily, the knowledge of the partition function is not necessary if
 one is interested to study the excluded volume interactions, which
 arise in the strong coupling limit. In fact, in this case
it is possible to apply a powerful method
due to Kleinert \cite{klI,klII,klbookII}.
This method
turns
the weak coupling expansion into
a strong coupling expansion which is convergent for large values of
$v_0$ and is able to accommodate also the anomalous dimensions of
quantum field theories.
The convergence of this strong coupling expansion is mostly very fast,
so that only a few coefficients of the weak coupling expansion 
must be known, see Refs.~\cite{kleinert,klbookII} for more details.  
These coefficients can be easily computed
starting from the well known
partition function of the Aharonov-Bohm
problem 
 without the
insertion of the $\delta-$function potential \cite{kleinert} and treating
this
potential as a small perturbation assuming that the
value of $v_0$ is small. The application of Kleinert's method in order
 to complete the brief analysis of the strong coupling limit made in
 this paper is work in progress.

\end{document}